\documentclass[a4paper,11pt,twoside]{article}
\usepackage[INRIA]{lipinria}
\usepackage{a4wide}

\usepackage{amsfonts}
\usepackage{amsmath}
\usepackage{amssymb}
\usepackage{amstext}
\usepackage{latexsym}
\usepackage{paralist}
\usepackage{url}
\usepackage{xspace}
\usepackage{subfigure}

\usepackage{amsthm}

\usepackage{figlatex}
\graphicspath{{fig/}}

\AtBeginDocument{}

\def\ie{i.e.,\xspace}

\newcommand{\calc}{w}    % calc for  stage
\newcommand{\comm}{\mathsf{\delta}} % comm for stage
\newcommand{\speed}{s}   % speed of proc
\newcommand{\bwprocp}{B}      % bandwidth of processor card
\newcommand{\bwlinkp}{b}      % bandwidth of link between 2 procs
\newcommand{\dbrate}{rate} % object rate
\newcommand{\dbsize}{d} % object size
\newcommand{\dbfreq}{f}    % object update frequency
\newcommand{\obj}{ob}
\newcommand{\op}{op}
\newcommand{\PP}{\mathcal{P}} % processors
\newcommand{\OO}{\mathcal{OB}} % objects
\newcommand{\OP}{\mathcal{OP}} % operators
\newcommand{\KK}{\mathcal{K}} % applications
\newcommand{\allocn}{a} %lloc}_{\textsf{n}}}
\newcommand{\allocop}{a_{op}} %lloc}_{\textsf{n}}}
\newcommand{\allocni}{\bar{a}} %lloc}_{\textsf{n}}}
\newcommand{\download}{Do}
\newcommand{\HET}{\textsc{Het}\xspace}
\newcommand{\HOM}{\textsc{Hom}\xspace}

\newcommand{\PROCNB}{\textsc{Proc-Nb}\xspace}
\newcommand{\PROCPW}{\textsc{Proc-Power}\xspace}
\newcommand{\BWSUM}{\textsc{BW-Sum}\xspace}
\newcommand{\BWMAX}{\textsc{BW-Max}\xspace}

\begin{document}
\RRInumber{2009-07}
\RRNo{6864}

\RRItitle{Resource Allocation for Multiple Concurrent In-Network Stream-Processing
  Applications}

\RRItitre{Stratégies d'allocation de ressources pour plusieurs applications
pipelinées concurrentes}

\RRIthead{Concurrent In-Network Stream Applications}
\RRIahead{A. Benoit \and H. Casanova \and 
   V. Rehn-Sonigo \and Y. Robert}

 \RRIauthor{Anne Benoit \and Henri Casanova \and 
   Veronika Rehn-Sonigo \and Yves Robert}

\RRIdate{February 2009}

\RRIkeywords{ in-network stream-processing, trees of operators,
  multiple concurrent applications, operator mapping, polynomial heuristics.} 

\RRImotscles{traitement de flux en réseau, arbres d'opérateurs,
multiples applications concurrentes, placement d'opérateurs, heuristiques polynomiales.}

\RRIabstract{
This paper investigates the operator mapping problem for in-network
stream-processing applications. In-network stream-processing amounts to
applying one or more trees of operators in steady-state, to multiple data
objects that are continuously updated at different locations in the
network. The goal is to compute some final data at some desired rate.
Different operator trees may share common subtrees.  Therefore, it may be
possible to reuse some intermediate results in different application trees.

The first contribution of this work is to provide complexity results for
different instances of the basic problem, as well as integer linear program
formulations of various problem instances.  The second second contribution
is the design of several polynomial-time heuristics. One of the primary
objectives of these heuristics is to reuse intermediate results shared by
multiple applications. Our quantitative comparisons of these heuristics
in simulation demonstrates the importance of choosing appropriate processors
for operator mapping. It also allow us to identify a heuristic that
achieves good results in practice.
}

\RRIresume{Dans ce rapport, on s'intéresse à des applications qui traitent des flux de données de manière pipelinée.
Chaque application consiste en un arbre d'opérateurs qu'on applique sur les données successives.
En régime permanent, les opérateurs interrogent des bases de données qui sont mises à jour périodiquement.
L'objectif est de calculer le résultat final à un débit fixé.
Des arbres d'opérateurs distincts peuvent partager des résultats.
De ce fait, il peut être possible de réutiliser quelques résultats intermédiaires pour
traiter différentes applications.

La première contribution de ce travail est l'obtention de résultats de complexité
pour les différentes instances du problème, ainsi que la formulation de ces
instances en terme de programme linéaire.
La deuxième contribution est le développement de plusieurs heuristiques polynomiales.
La réutilisation des résultats intermédiaires partagés par plusieurs applications
est un objectif premier de ces heuristiques. Nos comparaisons par simulation des heuristiques
démontrent toute l'importance du choix des processeurs pour le placement des opérateurs.
De même elles nous permettent d'identifier une heuristique performante, qui obtient de bons résultats
dans la pratique. }

\RRItheme{\THNum}

\RRIprojet{GRAAL}

\RRImaketitle

\section{Introduction}
\label{sec.intro}

% \remark{In this second paper, we concentrate on databases
% applications, in a non-constructive setting, and with servers which
% can compute. Thus we do not distinguish ``servers'' from
% ``processors''. The objective is mainly the bandwidth consumption of
% the network, though we could consider others. Complexity, ILP and
% heuristics need to be revisited.}

We consider the execution of applications structured as trees of operators,
where the leaves of the tree correspond to basic data objects that are
distributed over servers in a distributed network.  Each internal node in
the tree denotes the aggregation and combination of the data from its
children, which in turn generates new data that is used by the node's
parent. The computation is complete when all operators have been applied up
to the root node, thereby producing a final result.  We consider the
scenario in which the basic data objects are constantly being updated,
meaning that the tree of operators must be applied continuously. The goal
is to produce final results at some desired rate.  This problem is called
\emph{stream processing}~\cite{badcock_VLDB_2004} and arises in several
domains.

An important domain of application is the acquisition and refinement of
data from a set of sensors~\cite{srivastava_PODS2005, madden_ICMD_2003,
bonnet_CMDB_2001}.  For instance, \cite{srivastava_PODS2005} outlines a
video surveillance application in which the sensors are cameras
located at different locations over a geographical
area. The goal of the application could be to identify monitored areas
in which there is significant motion between frames, particular
lighting conditions, and correlations between the monitored
areas. This can be achieved by applying several operators (e.g.,
filters, pattern recognition) to the raw images, which are
produced/updated periodically.
Another example arises in the area of network
monitoring~\cite{cranor_ICMD_2002, vanRennesse_IPTPS_2002,
cooke_USENIX_2006}.  In this case routers
produce streams of data pertaining to forwarded packets.  More generally,
stream processing can be seen as the execution of one of more ``continuous
queries'' in the relational database sense of the term (e.g., a tree of
join and select operators). A continuous query is applied continuously,
i.e., at a reasonably fast rate, and returns results based on recent data
generated by the data streams. Many authors have studied the execution of
continuous queries on data streams~\cite{Babu_SIGMODRECORD_2001,
liu_tkde1999, chen02design, Plale_TPDD_2003, kramer_COMAD05}.

In practice, the execution of the operators must be distributed over the
network.  In some cases the servers that produce the basic objects may not
have the computational capability to apply all operators. Besides,
objects must be combined across devices,
thus requiring network communication.  Although a simple solution is to send
all basic objects to a central compute server, it often proves unscalable 
due to network bottlenecks. Also, this central server may
not be able to meet the desired target rate for producing results due to
the sheer amount of computation involved.  The alternative is then to distribute
the execution by mapping each node in the operator tree to one or more
servers in the network, including servers that produce and update basic
objects and/or servers that are only used for applying operators.  One then
talks of \emph{in-network stream-processing}. Several in-network
stream-processing
systems have been developed~\cite{abadi2005design,
medusa, pier, gates, nath-irisnet, vanRennesse_IPTPS_2002, NiagaraCQ,
MORTAR}. These systems all face the same question:
where should operators be mapped in the network?
%to which servers should one map which operators?

In this paper we address the operator-mapping problem for \emph{multiple
concurrent in-network stream-processing applications}. The problem for a
single application was studied in~\cite{Pietzuch_ICDE06} for an ad-hoc
objective function that trades off application delay and network bandwidth
consumption. In a recent paper~\cite{APDCM_09} we have studied a more
general objective function, enforcing the constraint that the rate at which
final results are produced, or \emph{throughput}, is above a given
threshold.  This corresponds to a Quality of Service (QoS) requirement of
the application and the objective is to meet this requirement while using
as few resources as possible.
In this paper we extend the work in~\cite{APDCM_09} in two ways.
First we study a ``non-constructive'' scenario, i.e., we are
given a set of compute and network elements, and we attempt to
use as few resources as possible while meeting QoS requirements. Instead,
in~\cite{APDCM_09}, we studied a ``constructive'' scenario in which resources
could be purchased and the objective was to spend as little money as possible.
Second, and more importantly, while in~\cite{APDCM_09} we studied the case of a single application,
in this paper we focus on multiple concurrent applications that
contend for the servers. Each application has its own QoS requirement and the
goal is to meet them all.
In this case, a clear opportunity for higher performance with
a reduced resource consumption is to reuse common sub-expression between
operator trees when applications share basic objects~\cite{Pandit06}.  We restrict our study to trees of
operators that are general binary trees and discuss relevant special cases
(e.g., left-deep trees~\cite{ioannidis96query}).  We consider target
platforms that are either fully homogeneous, or with a homogeneous network
but heterogeneous servers, or fully heterogeneous. Our specific contributions
are twofold:
%\begin{itemize}
%\item
(i) we formalize operator mapping problems for multiple in-network
      stream-processing applications and give their complexity;
(ii) we propose a number of algorithms to solve the problems and
      evaluate them via extensive simulation experiments. %;
%% (iii) we discuss in Appendix~\ref{sec.mutable}
%% how our results can be extended to the case of mutable
%%       applications, i.e., ones in which operators can be
%%       based on operator associativity and commutativity rules as in
%%       relational database applications~\cite{chen02design}.
%\end{itemize}
%

The rest of this paper is organized as follows. In
Section~\ref{sec.framework} we define our application and platform models,
and we formalize a number of operator mapping problems. In
Section~\ref{sec.comp} we discuss the computational complexity of our mapping
problems and in Section~\ref{sec.lp} we give integer linear programming
formulations.
In Section~\ref{sec.heur} we propose several heuristics, which we
evaluate in Section~\ref{sec.exp}. %Section~\ref{sec.mutable} discusses how our
%results can be extended to consider the case of mutable operator trees.
%We review related work in Section~\ref{sec.related} and
Finally we conclude
in Section~\ref{sec.conclusion} with a summary of our results and
future directions.

\section{Framework}
\label{sec.framework}

%We study operator mapping for multiple trees of operators which should
%be executed concurrently. First we describe the application and the
%platform, and then we introduce several relevant optimization problems.

\subsection{Application Model}
\label{sec.model.appli}

%$\NN=\{n_1, n_2, \dots\}$ is the global set of operators.

\begin{figure}[h]
   \centering
     \includegraphics[width=0.70\textwidth]{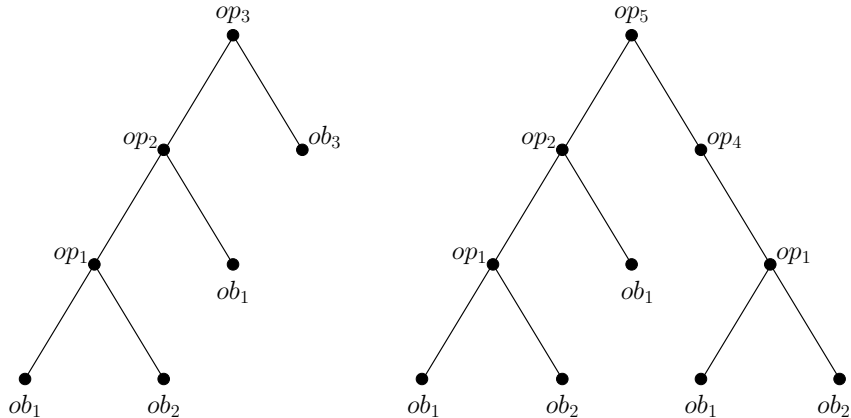}
   \caption{Sample applications structured as binary trees of operators.}
\label{fig.binarytree}
\end{figure}

We consider $\KK$ applications, each needing to perform several operations
organized as a binary tree (see Figure~\ref{fig.binarytree}). Operators are
taken from the set $\OP=\{\op_1, \op_2, \dots\}$, and operations are
initially performed on basic objects from the set $\OO=\{\obj_1, \obj_2,
\dots\}$. These basic objects are made available and continuously updated
at given locations in a distributed network.  Operators higher in the tree
rely on previously computed intermediate results, and they may also require
to download basic objects periodically.

%An operator $n_i$ may need some basic objects for computation and/or
%the result of computation of other operators.
%The leaves of each application trees are thus basic objects, and
%several leaves of the same tree or of different trees may
%correspond to the same object, as illustrated in the figure.

%Internal nodes (labeled $n_1^{(k)}, n_2^{(k)}, n_3^{(k)}, \dots$)
%represent operator computations for application~$k$. Several
%applications may perform the same operations, for instance
%$n_i^{(k)}=n_{i'}^{(k')}$.

For an operator $\op_p$ we define
%\begin{compactitem}
%\item
$objects(p)$ as the index set of the basic objects in $\OO$ that are needed
for the computation of $\op_p$, if any;
%\item
and $operators(p)$ as the index set of operators in $\OP$ whose
intermediate results are needed for the computation of $\op_p$, if
any.
%\item $\Parent^{(k)}(i)$: the index of the node's parent in $\NN^{(k)}$, if it exists.
%\end{compactitem}
We have the constraint that $|objects(p)| + |operators(p)| \leq 2$
since application trees are binary. An application is fully
defined by the operator at the root of its tree.
For instance, if we consider Fig.~\ref{fig.binarytree}, we have
one application rooted on~$\op_3$, and another application rooted
on~$\op_5$. Operator~$\op_1$ needs to download objects~$\obj_1$
and~$\obj_2$,  while operator~$\op_2$ downloads only object~$\obj_1$ but
also requires an intermediate result from operator~$\op_1$.

The tree structure of application~$k$ is defined with a set of
labeled nodes. The $i^{th}$ internal node in the tree of application~$k$
is denoted as $n_i^{(k)}$, its associated operator is
denoted as $op(n_i^{(k)})$, and the set of basic objects required by
this operator is denoted as $ob(n_i^{(k)})$.
%\begin{compactitem}
%\item
Node $n_1^{(k)}$ is the root node. %, which
%corresponds to operator $op(n_1^{(k)})$.
%\item
Let $\op_p=op(n_i^{(k)})$ be the operator associated to
node~$n_i^{(k)}$.
% If node $n_i^{(k)}$ has an associated operator
%$op(n_i^{(k)})=\op_p \in\OP$, then it must download basic objects in
%$ob(n_i^{(k)})=\{\obj_\ell\;|\;\ell\in objects(p) \}$.
Then node~$n_i^{(k)}$ has
$|operators(p)|$ child nodes, denoted as $n_{2i}^{(k)},n_{2i+1}^{(k)}$
if they exist.
%\item
Finally, the parent of a node $n_i^{(k)}$, for $i>1$, is the node of
index $\lfloor i/2 \rfloor$ in the same tree.
%\end{compactitem}

%All functions above are extended to sets of nodes:
%$f(I) = \cup_{i\in I} f(i)$, where $I$ is an index set and $f$ can be
%$\Leaf$ or $\Child$. % or $\Parent$.

The applications must be executed so that they produce final results,
where each result is generated by executing the whole operator tree
once, at a target rate. We call
this rate the application \emph{throughput}, $\rho^{(k)}$,
and the specification of the
target throughput is a QoS requirement for each application.
Each  %node $n_i^{(k)}$   %\in \NN^{(k)}$
operator in the tree of the $k^{th}$ application must compute
(intermediate) results at a rate at least as
high as the target application throughput~$\rho^{(k)}$.
Conceptually, operator $\op_p$ executes two concurrent threads in
steady-state:\\
 %\begin{itemize}
 %  \item
$\bullet$
It periodically downloads the most recent copies of the basic
   objects in $objects(p)$, if any.  %For our example
%   tree in Figure~\ref{fig.tree1}, $n_1$ needs to download $\obj_1$ and
%   $\obj_2$ while $n_2$ downloads only $\obj_1$ and $n_4$ does not
%   download any basic
%   object.
   Note that these downloads may simply amount to constant
   streaming of data from sources that generate data streams. Each
   download has a prescribed cost in terms of bandwidth based
   on application QoS requirements (e.g., so that computations are
   performed using sufficiently up-to-date data).  A basic object
   $\obj_j$ has a size
   $\dbsize_j$ (in bytes) and needs to be downloaded by the processors
   that use it for application~$k$ with frequency
   $\dbfreq_j^{(k)}$. Therefore,
   these basic object downloads consume an amount of bandwidth equal
   to $\dbrate_j^{(k)} =
   \dbsize_j \times \dbfreq_j^{(k)}$ on each network link and network card
   through which this object is communicated for
application~$k$. Note that if a processor requires object~$\obj_j$ for
several applications with different update frequencies, it downloads
the object only once at the maximum required frequency
$\dbrate_j=\max_k \{\dbrate_j^{(k)}\}$.\\  %Note that downloading $\obj_k$
%   to a location in the distributed network requires at least $\dbsize_k$
%   bytes of available disk space at that location.
 % \item
$\bullet$
It receives intermediate results computed by $operators(p)$,
    if any, and
    it performs some computation using basic objects it is
    continuously downloading, and/or
    data received from other operators.
    The operator produces some output, which is either an intermediate
result which will be sent to another operator, or the final result of
the application (root operator).
 The computation of operator
   $\op_p$ (to evaluate the operator once) requires
   $\calc_p$ operations, and
   produces an output of size $\comm_p$.

%\end{itemize}

%\begin{figure}
%  \begin{center}
%  \includegraphics[height=5cm]{leftdeeptree.fig}
%  \end{center}
%\caption{An example application structured as a left-deep tree of operators.}
%\label{fig.leftdeeptree}
%\end{figure}

%\remark{do we need left-deep trees in this setting??}
%In this paper we sometimes consider \emph{left-deep} trees, \ie binary
%trees in which the right child of an internal node is always a leaf. These
%trees arise in practical settings~\cite{XXX,XXX} and we show an
%example of left-deep tree in Figure~\ref{fig.leftdeeptree}.
%Here $\Child(i)$ and $\Leaf(i)$ have cardinal $1$ for every internal node
%$n_i$ but for the bottom-most internal node, $n_j$, for which
%$\Child(j)$ has cardinal $0$, and $\Leaf(j)$ has cardinal $1$ or $2$
%depending on the application.
%Note that the
%leftmost leaf $\inn$ is used only for symmetry, and represents an actual or
%fictitious flow of data.

%\begin{figure}
%\centering
%$$\begin{array}{cccccccccccccccccccccccccccccccccccccccc}
%\inn & \rightarrow & n_1 &  \rightarrow & n_2 & \rightarrow & n_3 & \rightarrow & n_4 & \rightarrow & n_5 & \rightarrow & n_6 & \rightarrow & \out\\
%& & \uparrow & & \uparrow & & \uparrow & & \uparrow & & \uparrow & & \uparrow\\
%& & A & & B & & A & & C & & B & & A
%\end{array}$$
%\caption{A left-deep application tree.}
%\label{fig.leftdeeptree}
%\end{figure}

\subsection{Platform Model}
\label{sec.model.platform}

The target distributed network is a fully connected graph (\ie a clique)
interconnecting a set of processors~$\PP$. These processors can
be assigned operators of the application tree and perform some
computation. Some processors also hold and update basic objects.
Each processor $P_u\in \PP$  is interconnected to the
network via a network card with maximum bandwidth  $\bwprocp_u$.
The network link between
two distinct processors $P_u$ and $P_v$ is
bidirectional and has bandwidth $\bwlinkp_{u,v}(=\bwlinkp_{v,u})$ shared
by communications in both directions.
In addition, each processor $P_u \in \PP$ is
characterized by a compute speed $\speed_u$.
% and a storage capacity $\procdisk_u$.
%
Resources operate under the full-overlap, bounded multi-port
model~\cite{bohong04}. In this model, a processor $P_u$ can be involved in
computing, sending data, and receiving data simultaneously. %Note that
%servers only send data, while processors engage in all three activities.
%Processor $P_u$, which is either a server or a processor, can be connected to multiple network links (since we assume a clique
%network).
The ``multi-port'' assumption states that each processor can send/receive
data simultaneously on multiple network links. The ``bounded'' assumption
states that the total transfer rate of data sent/received by processor $P_u$
is bounded by its network card bandwidth  $\bwprocp_u$.
The case in which some dedicated processors are only
providing basic objects but cannot be used for computations is
obtained simply by setting their compute speed to $0$.

\subsection{Mapping Model and Constraints}

Our objective is to map internal nodes of application
trees onto processors.  %If operator~$n_i$ appears in several nodes, it
%might be mapped onto distinct processors for each node.
As explained in Section~\ref{sec.model.appli}, if the operator
associated to a node requires basic objects, the processor in charge
of this internal node must continuously download up-to-date basic
objects, which consumes bandwidth on its processor's network card.  Each
used processor is in charge of one or several nodes.
If there is only
one node on processor $P_u$, while the processor computes for the
$t$-th final result it sends to its parent (if any) the data corresponding to
intermediate results for the $(t-1)$-th final result. It also receives data
from its children (if any) for computing the $(t+1)$-th final
result. All three activities are concurrent (see
Section~\ref{sec.model.platform}).
Note however that different nodes can be assigned to the same processor. In
this case, the same overlap happens, but possibly on different result
instances (an operator may be applied for computing the $t_1$-th result
while another is being applied for computing the $t_2$-th).  A particular
case is when several nodes with the same operator are assigned to the same
processor. In this case, computation is done only once for this operator,
but it should occur at the highest required rate among those of
the corresponding applications.

%Thus, the time required by each activity must be summed for all
%distinct operators to determine the processor's computation time.
%The \emph{cycle-time} of a processor is a maximum time taken by these
%three activities. The \emph{period} of the mapping is the largest
%cycle-time across the processors. The achieved throughput is the
%inverse of the period.

%\remark{and if we have different throughput for each app: the
%definition must be more elaborated, some operators going at different
%rates depending on the application... and for common operators between
%applications going at different rate? seems a bit tricky...}

A basic object can be duplicated, and thus available
and updated at multiple processors. We assume that duplication of
basic objects is
achieved in some out-of-band manner specific to the target application
(e.g., due to the use of a distributed database infrastructure that allows
consistent data replication).  In this case, a processor can choose among
multiple data sources when downloading a basic object, or perform a
local access if the basic object is available locally.
Conversely, if two nodes require the same basic object and are mapped to
different processors, they must both continuously download that object (and
incur the corresponding network overheads.)

We use an allocation function, $\allocn$, to denote the mapping of the nodes onto the
processors in~$\PP$: %, introducing
%the following notations:
%\begin{compactitem}
%\item
$\allocn(k,i) = u$ if node
$n_i^{(k)}$ is mapped to processor $P_u$. % (allocation function).
%\item
Conversely, $\allocni(u)$ is the index set of nodes mapped on $P_u$:
$\allocni(u) = \{(k,i)\;|\;\allocn(k,i)=u\}$.
%\item
Also, we denote by $\allocop(u)$ the index set of operators mapped on $P_u$:
 $\allocop(u) = \{p\;|\;\exists (k,i)\in \allocni(u) \;\;
 \op_p=op(n_i^{(k)})\}$. We introduce the following notations:
\begin{compactitem}
\item $Ch(u)=\{ (p,v,k) \}$ is the set of
(operator, processor, application) tuples such that
processor $P_u$ needs to receive an intermediate result
computed by operator~$\op_p$, which is mapped to processor~$P_v$, at rate $\rho^{(k)}$; operators $\op_p$ are
children of $\allocop(u)$ in the operator tree.
\item $Par(u)=\{ (p,v,k) \}$ is the set of (operator, processor,
application) tuples such that $P_u$ needs to send to $P_v$ an intermediate
result computed by operator~$\op_p$ at rate
$\rho^{(k)}$; $p\in\allocop(u)$ and
the sending is done to the parents of $\op_p$ in the operator tree.
\item $\download(u)= \{ (j,v,k) \}$ is the set of (object, processor, application) tuples
where $P_u$ downloads object $\obj_j$ from processor~$P_v$ at rate
$\rho^{(k)}$.
\end{compactitem}

\noindent The formal definition of $Ch(u)$ and $Par(u)$ is as follows. We first
define two sets of tuples, $ACh(u)$ and $APar(u)$, used to
account for communications for the same data but for different
applications:
{\footnotesize$$ACh(u) = \left\{ (p,v,k) \; | \; \exists i,p' \;\;\;
  p\in \allocop(v); p' \in \allocop(u); p\in operators(p');
  \op_p = op(n_i^{(k)}); \op_{p'} = op(n_{\lfloor i/2 \rfloor}^{(k)}
\right\}$$
$$APar(u) = \left\{ (p,v,k) \; | \; \exists i,p' \;\;\;
  p\in \allocop(v); p' \in \allocop(u); p\in operators(p');
  \op_p = op(n_i^{(k)}); \op_{p'} = op(n_{\lfloor i/2 \rfloor}^{(k)}
\right\}$$}
\noindent Then we determine which application has the higher
throughput for redundant entries, where $\arg\max$ randomly chooses
one application if there are equalities:
{\footnotesize$$kchosen(p,v,X)=\arg\max_{k\in\KK}\left\{  \rho^{(k)} \;|\; \exists
(p,v,k)\in X \right\}$$}
%The above $\arg\max$ chooses randomly one application if
%there are equalities.
\noindent Finally,
$X(u)=\{(p,v,kchosen(p,v,AX)) \;|\; op_p\in \OP, P_v \in \PP\}.$
%\noindent
$X$ stands for $Ch$ or $Par$, and we have thus
thus fully defined $Ch(u)$ and $Par(u)$.
\smallskip

Given these notations, we can now express constraints for the
application throughput: each processor must compute and
communicate fast enough to respect the prescribed throughput of each
application which is being processed by it.
The computation constraint is expressed below. Note that each operator
is computed only once at the maximum required throughput.

{\footnotesize
\begin{equation}
\label{eq.comp}
 \forall P_u\in\PP \;\;
\sum_{p\in \allocop(u)} \left(  \max_{(k,i)\in \allocni(u) \;|\;
op(n_i^{(k)})=\op_p} \left(\rho^{(k)} \right)
\frac{\calc_p}{\speed_u}\right) \;\; \leq 1 \;.
\end{equation}
}

%  $$\cycletime(u) = \max \left(
%\sum_{j \in \Child(\allocni(u))\setminus \allocni(u)} \frac{\comm_j}{\bwlinkp_{\allocn(j),u}},~~
%       \sum_{i\in \allocni(u)} \frac{\calc_i}{\speed_{u}},~~
%       \sum_{j \in \Parent(\allocni(u))\setminus \allocni(u)~}
%\sum_{i\in \Child(j)\cap \allocni(u)} \frac{\comm_i}{\bwlinkp_{u,\allocn(j)}} \right)\;$$

Communication occurs only when a child or the parent of a given
node and this node are mapped on different processors. In other terms,
we neglect intra-processor communications.
An operator computing for several applications may send/receive results
to/from different processors. If the parent/child nodes corresponding
to the different applications are mapped onto the same processor, the
communication is done only once, at the most constrained throughput.
This throughput, as well as the processors with which $P_u$ needs to
communicate, are obtained via $Ch(u)$ and $Par(u)$. In these
expressions $v\neq u$ since we neglect intra-processor-communications.

The first part of Eq.~\ref{eq.com.rec} expresses constraints on
receiving, %while
%Eq.~\ref{eq.com.send}
while the second part refers to sending:
{\footnotesize
\begin{equation}
\label{eq.com.rec}
 \forall P_u\in\PP \;\;  \sum_{(p,v,k)\in Ch(u)} \left( \rho^{(k)}
\frac{\comm_p}{\bwlinkp_{v,u}} \right) \;\; \leq 1 \; ; \quad
%\end{equation}
%
%\begin{equation}
%\label{eq.com.send}
 \forall P_u\in\PP \;\;  \sum_{(p,v,k)\in Par(u)} \left( \rho^{(k)}
\frac{\comm_p}{\bwlinkp_{u,v}} \right) \;\; \leq 1\;.
\end{equation}
}
%\remark{well, for the sending, needs to keep track of ``i'' too, so
%it's gonna be a bigger mess... still should formalize Ch(u) and Par(u)
%too...}

%The period of the mapping is defined as the largest cycle-time:
%$\period = \max_{P_u \in \PP} \cycletime(u)$,
%with the overall achieved throughput equal to the inverse of $\period$.
%A processor not used in the computation has a cycle-time of~$0$.

$P_u$ must have enough bandwidth capacity to perform all its basic
object downloads, to support downloads of the basic objects it
may hold, and also to perform all communication with
other processors, all at the required rates.
This is expressed in Eq.~\ref{eq.bwproc}.
The first term corresponds to basic object downloads; the second term
corresponds to download of basic objects from other processors;
 the third term corresponds to inter-node communications when a node
is assigned to $P_u$ and its parent node is assigned to another
processor; and the last term corresponds to inter-node communications
when a node is assigned to $P_u$ and some of its children  nodes are
assigned to another processor.
{\footnotesize\begin{equation}
\label{eq.bwproc}
 \forall P_u\in \PP \;\;\\
  \sum_{(j,v,k) \in \download(u)} \dbrate_j^{(k)} +
  \sum_{P_v\in \PP} \; \sum_{(j,u,k)\in \download(v)} \dbrate_j^{(k)} +
  \sum_{(p,v,k)\in Ch(u)} \delta_p \rho^{(k)}  +
  \sum_{(p,v,k) \in Par(u)} \delta_p \rho^{(k)}
     \leq \bwprocp_u
\end{equation}}

Finally, we need to express the fact that the link
 between processor $P_u$ and processor $P_v$ must have
  enough bandwidth capacity to support all possible communications between
  the nodes mapped on both processors, as well as the object downloads
between these processors. Eq.~\ref{eq.bwlink} is similar to
Eq.~\ref{eq.bwproc}, but it considers two specific processors:
{\footnotesize\begin{equation}
\label{eq.bwlink}
 \forall P_u,P_v\in \PP \;\; \\
  \sum_{(j,v,k) \in \download(u)} \dbrate_j^{(k)} +
  \sum_{(j,u,k) \in \download(v)} \dbrate_j^{(k)} +
  \sum_{(p,v,k)\in Ch(u)} \delta_p \rho^{(k)}  +
  \sum_{(p,v,k) \in Par(u)} \delta_p \rho^{(k)}
     \leq \bwlinkp_{u,v}
\end{equation}}

\subsection{Optimization Problems}
\label{sec.prob.obj}

The overall objective of the operator-mapping problem is to ensure that a
prescribed throughput per application is achieved while minimizing a
cost function. Several relevant problems can be envisioned.
%
%\begin{compactitem}
%\item
\PROCNB minimizes the number of processors enrolled for computations
(processors that are allocated at least one node);
\PROCPW minimizes the compute capacity and/or the network
card capacity of processors enrolled for computations (e.g., a
linear function of both criteria);
\BWSUM minimizes the sum of the bandwidth capacities used by the application;
and finally \BWMAX minimizes the maximum percentage of bandwidth used on all
links (minimizing the impact of the applications on the network for
other users).
%\end{compactitem}
%\medskip

Different platform types may be considered depending on the heterogeneity
of the resources.  We consider the case in which the platform is fully
homogeneous ($\speed_u=\speed$, $\bwprocp_u = \bwprocp$ and $\bwlinkp_{u,v}
= \bwlinkp$), which we term \HOM. The
 %We then consider the case in which the
%processors are heterogeneous but the network links are homogeneous
%($\bwlinkp_{u,v} = \bwlinkp$), which we term \LAN.  Finally we consider the
heterogeneous case in which network links can have various
bandwidths is termed \HET.

Each combination of problems and platforms could be envisioned, but we will
see that \PROCPW on a \HOM platform is actually equivalent to \PROCNB.
\PROCNB makes more sense in this setting, while \PROCPW is used for
%\LAN and
\HET platforms only. Both types of platforms are considered for
the \BWSUM and \BWMAX problems.

%\remark{Idea: We only deal with achieving a threshold throughput at minimum "cost".
%This notion of cost for now is simply the number of processors needed.
%Other ideas that we may consider, although probably not, are some notion of
%overall number of CPU cycles used per second, and over number of bytes
%transfered by second}

%\remark{Note: if you really want to maximize the throughput without going over a prescribed cost, then you can simply use a binary search and solve the previous problem repeatedly}

\section{Complexity}
\label{sec.comp}

Problem \PROCNB is NP-complete in the strong sense. This is true even for a
simple case: a \HOM platform and a single application ($|\KK| = 1$), that
is structured as a left-deep tree, in which all operators take the same
amount of time to compute and produce results of size 0, and in which all
basic objects have the same size.  We refer the reader to a technical
report for the proof~\cite{RR2008-20}, which relies on a straightforward
reduction to 3-Partition, which is known to be NP-hard in the strong
sense~\cite{GareyJohnson}.  It turns out that the same proof holds for
\PROCPW on a \HOM platform.
%Both problems become polynomial if one adds the
%restriction that no basic object is used by more than one operator in the
%application tree. In this case one can assign operators in round-robin
%fashion to arbitrary processors, simply picking the smallest number of
%processors necessary to respect bandwidth and computational power
%constraints.

The \BWMAX problem is NP-hard because downloading objects with different
rates on two processors is the same problem as 2-Partition, which is known
to be NP-hard~\cite{GareyJohnson}.  Here is a sketch of the straightforward
proof, which holds even in the case of a single application.  Consider an
application in which all operators produce zero-size results, and in which
each basic object is used only by one operator.  Consider three processors,
with one of them holding all basic objects but unable to compute any
operator.
%and each of the other two able to compute only half of the
%operators.
The two remaining processors are able to compute all the operators,
and they are connected to the first one with
identical network links. Such an instance can be easily constructed. The
problem is then to partition the set of operators in two subsets so that
the bandwidth consumption on the two network links in use is as equal as
possible. This is exactly the 2-Partition problem.

The \BWSUM problem is NP-hard because it can be reduced to the Knapsack
problem, which is NP-hard~\cite{GareyJohnson}. Here is a proof sketch for
a single application. Consider the same application as for the proof of the
NP-hardness of \BWMAX above. Consider two identical processors, $A$ and
$B$, with $A$ holding all basic objects. Not all operators can be executed
on $A$ and a subset of them need to be executed on $B$. Such an instance
can be easily constructed. The problem is then to determine the subset of
operators that should be executed on $A$. This subset should satisfy the
constraint that the computational capacity of $A$ is not exceeded, while
maximizing the bandwidth cost of the basic objects associated to the
operators in the subset. This is exactly the Knapsack problem.

All these problems can be solved thanks to an integer linear program
(see Section~\ref{sec.ilp}). However, they cannot
be solved in polynomial time (unless P=NP). Therefore, in the Section~\ref{sec.heur}
we describe polynomial-time heuristics for one of these problems.

\section{Linear Programming Formulation}
\label{sec.ilp}

In this section, we give an integer linear program (ILP) formulation of the
\PROCPW-\HET, \BWSUM-\HET and \BWMAX-\HET problems,  in terms of an integer
linear program (ILP). These are the most general versions of our
operator-mapping problems. More restricted versions, e.g., with \HOM
platforms, can be solved using the same ILPs. We describe the input data to
the ILP, its variables, its constraints, and finally its objective
functions.

In all that follows, $i$ and $i'$ are indices spanning nodes in set of
nodes of an application tree; $p$ and $p'$ are indices spanning operators
in $\OP$; $j$ is an index spanning objects in $\OO$; $u$, $u'$, and $v$ are
indices spanning processors in $\PP$; $k$ is an application index
spanning~$\KK$.
%between $1$ and $\KK$.

\subsection{Input Data}

Parameters $\comm_i, \calc_i$ for operators, $\dbrate_j^{(k)}$ for object
download rates, and $\speed_u, \bwprocp_u, \bwlinkp_{u,v}$ for processors
and network elements, are rational numbers and defined in
Section~\ref{sec.framework}.  $\rho^{(k)}$ is a rational number that
represents the throughput QoS requirement for application~$k$.  For
convenience, we also introduce families of boolean parameters: $par$, $oper$,
and $object$, that pertain to application trees; and $obj$, that pertain to
location of objects on processors. We define these parameters hereafter:

\begin{itemize}
\item $par(k,i,i')$ is equal to $1$ if internal node
$n_i^{(k)}$ is the parent of $n_{i'}^{(k)}$ in the tree of application~$k$,
and $0$ otherwise.

\item $oper(k,i,p)$ is equal to $1$ if $op(n_i^{(k)})=p$, and $0$ otherwise.

\item $object(k,i,j)$ is equal to $1$ if node $n_i^{(k)}$ needs object
$\obj_j$ (i.e., $p\in objects(op(n_i^{(k)}))$), and $0$
otherwise.

\item $obj(u,j)$ is equal to $1$ if processor $P_u$ owns a copy of
object~$\obj_j$, and $0$ otherwise.

\end{itemize}

\subsection{Variables}

\begin{itemize}
\item $x_{k,i,u}$ is a variable equal to $1$ if node $n_i^{(k)}$ is
mapped on $P_u$, and $0$ otherwise.

\item $d_{j,u,v,k}$ is a variable equal to $1$ if processor $P_u$
downloads object $\obj_j$ for application~$k$ from processor~$P_v$, and $0$
otherwise.

\item $y_{k,i,u,i',u'}$ is a variable equal to $1$ if $n_i^{(k)}$
is mapped on $P_u$, $n_{i'}^{(k)}$ is mapped on
$P_{u'}$, and $n_i^{(k)}$ is the parent of $n_{i'}^{(k)}$
in the application tree.
%There are $|\NN|^4.|\CC|^2$ such variables.

\item $used_u$ is a variable equal to $1$ if there is at
least one node mapped to processor $P_u$, and $0$ otherwise.

%%-----------------------------------------------------
\item $xop_{k,p,u}$ is a variable equal to $1$ if $op_p$ of
  application $k$ is mapped to processor $P_u$, and $0$ otherwise.

\item $yop_{k,p,u,p',u'}$ is a variable equal to $1$ if
  $op_p$ of application $k$ is mapped on processor $P_u$, $op_{p'}$ of
  application $k$ is mapped on processor $P_{u'}$, and $op_p$ is a parent of
  $op_{p'}$ in application $k$, and $0$ otherwise.

\item $Ch_{u,p,v,k}$ is a variable equal to $1$ if $(p,v,k)
  \in Ch(u)$, and $0$ otherwise.

\item $Par_{u,p,v,k}$ is a variable equal to $1$ if $(p,v,k) \in
  Par(u)$, and $0$ otherwise.

\item $rho_{u,p}$ is a rational variable equal to the
  throughput of $op_p$ if it is mapped on processor $P_u$, and $0$ otherwise.

\item $ratemax_{j,u,v}$ is a rational variable equal to the download rate
  of object $ob_j$ by processor $P_u$ from processor $P_v$, and $0$ otherwise.

\end{itemize}

\subsection{Constraints}

We first give constraints to guarantee that the allocation of nodes to
processors is valid, and that each required download is done from a server
that holds the relevant object.

\begin{itemize}

\item $\forall k,i~~ \sum\limits_{u} x_{k,i,u}=1$: each node is
placed on exactly one processor;

\item $\forall j,u,v,k~~ d_{j,u,v,k} \leq obj(v,j)$: object~$\obj_j$
can be downloaded from processor $P_v$ only if $P_v$ holds it;

\item
$\forall i,j,u,k~~ 1\geq \sum\limits_v d_{j,u,v,k} \geq
x_{k,i,u}.object(k,i,j)$: processor $P_{u}$ must download object~$\obj_j$
from exactly one processor $P_v$ if there is a node $n_i^{(k)}$ mapped on
processor $P_u$ that requires $\obj_j$.

%% \item $\forall i,k,c,u~~ 1\geq \sum_l d_{c,u,k,l} \geq
%% x_{i,c,u}.leaf(i,k)$: processor $P_{c,u}$ must download
%% object~$\obj_k$ from exactly one server if there is a node $n_i$
%% mapped on this processor that requires $\obj_k$ for computation.
\end{itemize}

\noindent
The next two constraints aim at properly defining
variables~$y$. Note that a straightforward definition would be
$y_{k,i,u,i',u'}=par(k,i,i').x_{k,i,u}.x_{k,i',u'}$, but this leads
to a non-linear program. Instead we write, for all $k,i,u,i',u'$:
%% Notice that a straightforward definition would be
%% $y_{i,c,u,i',c',u'}=par(i,j).x_{i,c,u}.x_{i',c',u'}$, but this leads
%% to a non-linear program. Instead we write (for all $i,c,u,i',c',u'$):
\begin{itemize}
\item $y_{k,i,u,i',u'} \leq par(k,i,i')$; $~y_{k,i,u,i',u'} \leq
x_{k,i,u}$; $~y_{k,i,u,i',u'} \leq x_{k,i',u'}$: $y_{k,i,u,i',u'}$ is forced to
be $0$ if one of these three conditions does not hold.
\item $y_{k,i,u,i',u'} \geq par(k,i,j).\left( x_{k,i,u} + x_{k,i',u'}-1
\right)$: $y_{k,i,u,i',u'}$ is forced to be $1$ only if the three
conditions are true (otherwise the right term is lower than or equal
to~$0$).

\end{itemize}

\noindent
The following two constraints ensure that $used_{u}$ is properly defined:
\begin{itemize}
\item $\forall u~ used_{u} \leq \sum\limits_{k,i} x_{k,i,u}$:
processor $P_{u}$ is not used if no node is mapped to it;
\item $\forall k,i,u~ used_{u} \geq x_{k,i,u}$: processor $P_{u}$ is
used if at least one node~$n_i$ is mapped to it.
\end{itemize}

\noindent
The following four constraints ensure that $xop_{k,p,u}$ and
$yop_{k,p,u}$ are properly defined:
\begin{itemize}
\item $\forall i,k,p,u~ xop_{k,p,u} \geq x_{k,i,u}.oper(k,i,p)$:
$xop$ is forced to be $1$ if operator $op_p$ of application $k$ is
  mapped on processor $P_u$;
\item  $\forall k,p,u~~ xop_{k,p,u} \leq \sum \limits_i x_{k,i,u}.oper(k,i,p)$: $xop$
  is forced to be $0$ if operator $op_p$ of application $k$ is not
  mapped on processor $P_u$;
\end{itemize}
\begin{itemize}
\item $\forall k,p,p',u,u',i,i'~~ yop_{k,p,u,p',u'} \leq xop_{k,p,u}$; $yop_{k,p,u,p',u'} \leq
  xop_{k,p',u'}$; $yop_{k,p,u,p',u'} \leq par(k,i,i')$;
  $yop_{k,p,u,p',u'} \leq oper(k,i,p)$; $yop_{k,p,u,p',u'} \leq
  oper(k,i',p')$: $yop_{k,p,u,p',u'}$ is forced to be $0$ if one of these conditions
  does not hold;
\item $\forall k,p,p',u,u',i,i'~~ yop_{k,p,u,p',u'} \geq
  par(k,i,i').oper(k,i,p).oper(k.i',p').(xop_{k,p,u} + xop_{k,p',u'} -
  1)$: $yop_{k,p,u,p',u'}$ is forced to be $1$ only if all five conditions are true.
\end{itemize}

\noindent
The next four constraints ensure that $Ch_{u,p,v,k}$ and $Par_{u,p,v,k}$ are
defined properly:
\begin{itemize}
\item $\forall u,p,v,k~~ Ch_{u,p,v,k} \leq \sum\limits_{p'}yop_{k,p',u,p,v}$: in
  application $k$, if the
  parent operator of operator $op_p$, which is mapped on $P_v$, is not mapped
  on processor $P_u$, $Ch$ is forced to be $0$;

\item $\forall p',u,p,v,k~~ Ch_{u,p,v,k} \geq yop_{k,p',u,p,v}$: in
  application $k$, if operator $op_p$ of application $k$ is mapped to
  $P_v$ and its parent operator in the application tree is mapped to $P_u$,
  $Ch$ is forced to be $1$.

\item $\forall u,p,v,k~~ Par_{u,p,v,k} \leq \sum\limits_{p'}yop_{k,p',v,p,u}$: in
  application $k$, if the
  parent operator of operator $op_p$, which is mapped to $P_u$, is not mapped
  to processor $P_v$, $Par$ is forced to be $0$;

\item $\forall p',u,p,v,k~~ Par_{u,p,v,k} \geq yop_{k,p',v,p,u}$:
  in application $k$, if operator $op_p$ is mapped to $P_u$ and its parent
  operator in the application tree is mapped to $P_v$, $Par$ is forced to
  be $1$.

\end{itemize}

\noindent
The following two constraints ensure that the throughput QoS requirement of each
application, $\rho^{(k)}$, is met:

\begin{itemize}
\item $\forall k,u,p~~rho_{u,p} \geq xop_{k,p,u}.\rho^{(k)}$: the
  throughput of processor $P_u$, to which operator $op_p$ of application
  $k$ is mapped, has to satisfy the throughput QoS requirement of
  application $k$;

\item $\forall k,p,u,v~~ratemax_{p,u,v} \geq
  d_{p,u,v,k}.rate_p^{(k)}$:
  the update rate of operator $op_p$ on processor $P_u$ has to satisfy the
  throughput QoS requirement of application $k$;

\end{itemize}

\noindent
The following constraint ensures that the compute capacity of
each processor is not exceeded while meeting QoS throughput
requirements:
\begin{itemize}
\item $\forall u~~\sum\limits_p rho_{u,p} \frac{\calc_p}{\speed_u}\leq 1$.
\end{itemize}

\noindent
The following two constraints ensure that the bandwidth capacity of network
elements are not exceeded:
\begin{itemize}
\item  Bandwidth constraint for the processor network cards:
%\allowdisplaybreaks
  \begin{equation}
    \begin{split}\forall u~~ & \sum\limits_{p,v,k} Ch_{u,p,v,k}.rho_{u,p}.\delta_p +
      \sum\limits_{p,v,k} Par_{u,p,v,k}.rho_{u,p}.\delta_p +\\
      & \sum\limits_{j,v,k} d_{j,u,v,k}.ratemax_{j,u,v}+
      \sum\limits_{j,v,k}
    d_{j,v,u,k}.ratemax_{j,v,u} \leq \bwprocp_u
    \end{split}
\end{equation}
\item Bandwidth constraints for links between processors:
  \begin{eqnarray}
    \begin{split}\forall u,v~~ &\sum\limits_{p,k} Ch_{u,p,v,k}.rho_{u,p}.\delta_p +
\sum\limits_{p,k} Par_{u,p,v,k}.rho_{u,p}.\delta_p +\\
&\sum\limits_{j,k} d_{j,u,v,k}.ratemax_{j,u,v}+
\sum\limits_{j,k}
d_{j,v,u,k}.ratemax_{j,v,u} \leq \bwlinkp_{u,v}
\end{split}
\end{eqnarray}

\end{itemize}

\subsection{Objective Function}

We have to define the objective function to optimize. We have a different
definition for each problem:
\medskip
%\paragraph{

\PROCPW-\HET:
\begin{equation}
\min\left( \sum\limits_{u,p} rho_{u,p}\frac{\calc_p}{\speed_u} \right) \;.
\end{equation}

%\paragraph{
\BWSUM-\HET:
  \begin{eqnarray}
   \begin{split}
     \min &\sum\limits_{u,v,p,k} Ch_{u,p,v,k}.rho_{u,p}.\delta_p +
\sum\limits_{u,v,p,k} Par_{u,p,v,k}.rho_{u,p}.\delta_p +\\
&\sum\limits_{u,v,j,k} d_{j,u,v,k}.ratemax_{j,u,v}+
\sum\limits_{u,v,j,k} d_{j,v,u,k}.ratemax_{j,v,u}  \;.
\end{split}
\end{eqnarray}

%\paragraph{
\BWMAX-\HET:  %}~\\
For this problem we need to add one variable, $bwmax$, and $|\PP|^2$ constraints:
  \begin{eqnarray}
    \begin{split}\forall u,v~~ &\sum\limits_{p,k} Ch_{u,p,v,k}.rho_{u,p}.\delta_p +
\sum\limits_{p,k} Par_{u,p,v,k}.rho_{u,p}.\delta_p +\\
&\sum\limits_{j,k} d_{j,u,v,k}.ratemax_{j,u,v}+
\sum\limits_{j,k}
d_{j,v,u,k}.ratemax_{j,v,u} \leq bwmax
\end{split}
\end{eqnarray}
\noindent
and the objective becomes: $\min\left( bwmax  \right)$.

\section{Heuristics}
\label{sec.heur}
In this section we propose several polynomial heuristics\footnote{To
  ensure the reproducibility of our results,
the code for all heuristics is available on the
web~\cite{code.multiApp}.}  for the
\PROCPW problem, in which we consider only the compute capacity of
processors enrolled for computation.
%As for processing the
%application nodes,
Two heuristics use a random approach to process application nodes,
while the others are based on tree traversals. As for the choice of an
appropriate resource for the current node,
four different processor selection strategies are implemented (and
shared by all heuristics).
Two selection strategies are \emph{blocking} and two are \emph{non-blocking}.
\emph{Blocking} means that once chosen for a given operator $op_1$,
a processor cannot be reused later for another operator $op_2$,
%%Vero: what are the "relatives" in the following line? what do you mean? children? suppress it?
and it is only possible to add relatives (i.e., father or children) of
$op_1$ to this processor.
%%Vero
%%Yves, je voulais dire qu'on ne peut ajouter que les voisins, donc pere
%%ou enfants, au proc
On the contrary, \emph{non-blocking}
strategies impose no such restrictions.
We start with a description of the %{\bf four %processor allocation
%selection strategies},
four processor selection strategies, and then we move to a brief overview of each
heuristic.

\subsection*{Processor Allocation Strategies}

\paragraph{(1)~Fastest processor first (blocking) --}
Every time we have to chose a processor,
the fastest remaining (not already chosen) processor is chosen.

\paragraph{(2)~Biggest network card first (blocking) --}
Every time we have to chose a processor, the remaining processor with
the biggest network card is chosen.

\paragraph{(3)~Fastest remaining processor (non-blocking) --}
The actual amount of computation is subtracted from the
computation capability, and the processor with the most remaining
computation power is chosen.

\paragraph{(4)~Biggest remaining network card (non-blocking) --}
In this strategy the current (already assigned) communication volume is
subtracted from the network card capacity to evaluate the processor
whose remaining communication capacity is the biggest. This processor
is chosen.

\subsection*{Significance of Node Reuse}
\begin{figure}[h]
  \centering
  \includegraphics[width=0.7\textwidth]{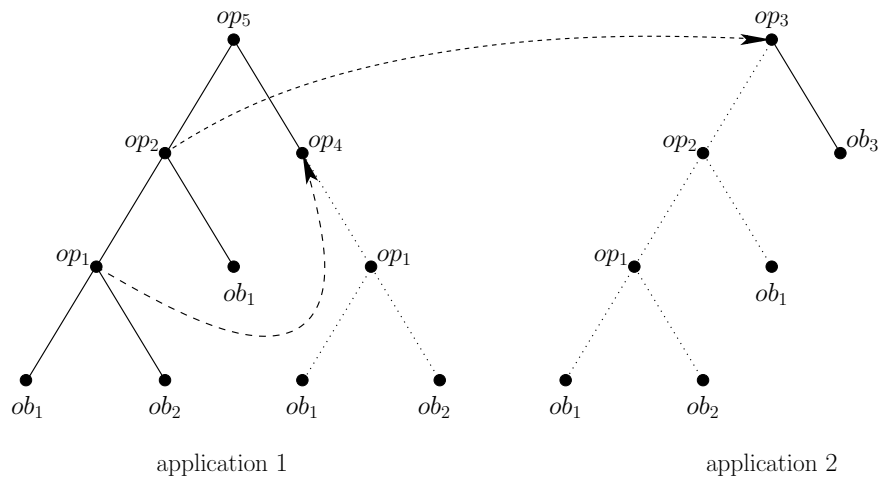}
  \caption{Example for the reuse of nodes. $op_1$ is only computed
    once and its result is reused for the computation of $op_2$ and
    $op_4$. $op_3$ uses the result of $op_2$ in application $1$ for
    its computation.}
  \label{fig.reuse}
\end{figure}

Our heuristics, except RandomNoReuse (H1),
%see Section~\ref{sec.randNo})
are designed for node reuse.
This means that we try to benefit from the fact that different
applications may have
common subtrees, i.e., subtrees composed of the same
operators. Instead of recomputing the result for such a subtree, we aim
at reusing the result. For this purpose we try to
add additional communications as can be seen in
Figure~\ref{fig.reuse}. The processor that computes the left $op_1$ in
application $1$ sends its result not only to the processor that computes
$op_2$, but also to the processor that computes $op_4$. The operator
$op_1$ on the right of application $1$ no longer has to be computed.
In the same way, we save the whole computation of the subtree
rooted by $op_2$ in application $2$ when we add the communication
between $op_2$ in application $1$ and $op_3$ in application $2$.

\medskip
\noindent
We give hereafter a brief overview of each heuristic:
\subsection*{H1: RandomNoReuse}
%\label{sec.randNo}
The H1 heuristic does not reuse any
result. While there are unassigned operators, H1 randomly
picks one of them. If the father is already mapped, it tries to map
the operator on the father's processor, or it tries the children's
processors, if those are already mapped. If none of these mappings is
possible, H1 chooses a new processor according to the
processor selection
strategy, and maps the operator. If this is not
possible, H1 fails.

\subsection*{H2: Random}
The H2 random heuristic is more sophisticated as it tries to reuse
common results. If the randomly chosen operator has not already been mapped,
possibly for another application, we use the same mechanism as in~H1: first
try to map the operator on its father's
processor or one of the children's, and in case of failure choose a
new processor. But, if the operator has already been mapped somewhere
else in the forest, we try to
add a link from the already mapped operator to the father of the
actual operator to reuse the common result. When this is possible, we mark
the whole subtree (rooted at the operator) as
mapped. Otherwise, we choose a new processor.

\subsection*{H3: TopDownBFS}
The H3 heuristic performs a breadth-first-search (BFS) traversal of all
applications. We use an artificial root node to link all
applications, i.e., all application roots become children of the
artificial root.
% to be able to make the BFS traversal over all applications.
For each operator, we check whether the operator has not been mapped yet
and whether its father has.
In this case, H3 tries to map the operator on the same
processor as its father, and in case of success continues the BFS
traversal. In the case where the actual operator has already been mapped onto
one or more processors, H3 tries to add a communication link
between the mapped operator and the father of the actual
operator: the mapped operator sends its result not
only to its father but also to the father of the actual operator.
If none of these two conditions holds, or if the mapping was not
possible, H3 tries to map the operator onto a new
processor. The processor is chosen according to the processor
selection strategy. When the mapping is successful, the BFS traversal
is continued, otherwise H3 fails.

\subsection*{H4: TopDownDFS}
The H4 heuristic uses the same mechanism as H3, but
operators are treated in depth-first-search (DFS)
manner. Thus, each time a mapping of a node is successful, the
heuristic continues the DFS traversal of the current application tree.

\subsection*{H5: BottomUpBFS}
As the H4 heuristic, the H5 heuristic makes a BFS traversal of the
application forest. For this purpose we use the same mechanism of
a new artificial root that links all applications. For each operator,
H5 verifies whether it has already been mapped
on a processor. In this case a communication link is added (if possible),
connecting the mapped operator and the father of the unmapped
operator. If the operator is not yet mapped and if it has some children, we
try to map the operator to one of its children's processors. If no such
possibility is successful, or if the operator is at the bottom
of a tree,
H5 tries to map the operator onto a new processor (where
the processor is chosen according to the processor selection
strategy).
When the mapping is successful, the BFS traversal continues,
otherwise H5 fails.
%If the mapping is not possible at this point, the heuristic fails.
%Every time an operator is mapped, we continue the BFS traversal.

\subsection*{H6: BottomUpDFS}
The H6 heuristic is similar to H5, but instead of a BFS traversal,
it performs a DFS traversal of the application forest. This
makes the heuristic a little bit more complicated, as there are more
cases to be considered.
For each node we check if its operator has already been mapped on
a processor, and none of its children are. In
this case we go up in the tree until we reach the last node $n_1$ such
that there exists a node~$n_2$ somewhere else in the forest which is
already mapped, and such that $op(n_1)=op(n_2)$.
% operator
%$op_{p}$ that was also associated to another node, $n_i^{(k)}$,
%somewhere else in the
%forest, which was already mapped.
%(i.e., an operator
%$op_{i2}$. $op_{i1}$ and $op_{i2}$ are the same type of operator
%at different locations in the forest).
%%Vero pas super clair; i1 = i2?
%%Yves mieux?
%%AB j'ai modifie, c'est qu'il faut parler de nodes
In this case we try
to add a communication between $n_2$
%this operator $op_{i2}$%
and the father of $n_1$ to benefit from the calculated result. If the children
have already been mapped we simply try to map the operator to one of
the children's processors. If this is not possible or if the additional
communication was not possible or again if the operator has not been mapped
anywhere in the forest, H6 tries to map the operator onto a
new processor, according to the processor selection strategy.
%When no appropriate processor can be found,
Otherwise H6 fails.

%%Vero je me rends compte
%%il y a une confusion entre reusing a node common to several pps and reusing a proc.
%% je remonte changer

%% \vspace{-0.2cm}
\section{Experimental Results}
\label{sec.exp}

We have conducted several experiments to assess the performance of the
different heuristics described in Section~\ref{sec.heur}. In particular, we
are interested in the impact of node reuse on the number of solutions found
by the heuristics.
%%Vero; c'est bien node reuse? ou operator-reuse?
%We outline the experimental plan in
%Section~\ref{sec.plan}. Sections~\ref{sec.numApp}
%to~\ref{sec.similarity} show results of several experimentation sets,
%where parameters like number of applications or number of available
%processors vary.
%
\subsection{Experimental Plan}
\label{sec.plan}
%We use $50$ randomly generated application trees of at
%most $N$ operators (non-leaf nodes in the trees).
Except for Experiment~1, all application trees are fixed to a size of
at most $50$ operators, and except for Experiment~5, we consider $5$
concurrent applications.
The leaves in the tree correspond to basic
objects, and each basic object is chosen randomly among 10 different
types. The size $\dbsize$ of each object type is also chosen randomly and varies
between 3MB and 13MB. The download frequencies of objects for each
application, $f$, % ($f_j^{(k)}$),
as well as the application throughput, $\rho$,
are chosen randomly such that $0<f\leq 1$ and
  % $0< f_j^{(k)}
%\leq 1$ and
$1\leq \rho \leq 2$.
%Recall that the download rate for object $o_k$ is then computed as
%$rate_j^{(k)}=\dbsize_j \times f_j^{(k)}$ for application $k$.
%
The parameters for operators are also
chosen randomly. In all experiments (except Experiment~4),
%, except for experiences concerning
%the communication-to-computation ratio (see Section~\ref{sec.CCR}),
the computation amount $w_i$ for an operator lies between $0.5$MFlop/sec and
$1.5$MFlop/sec, and the output size of each operator $\delta_i$ is randomly
chosen between $0.5$MB and $1.5$MB.

Throughout most of our experiences we use the following platform
configuration (variants will be mentioned explicitly when needed.)
We dispose of $30$ processors. Each processor is
equipped with a network card, whose bandwidth limitation varies between
$50$MB and $180$MB. We use the same range for computation power, i.e., CPU
speeds of $50$MIPS to $180$MIPS. The different processors are interconnected via
heterogeneous communication links, whose bandwidth are between
$60$MB/s and $100$MB/s.
The $10$ different types of objects are randomly distributed over the
processors.
Execution time and communication time are scaled units, thus
execution time is the ratio between computation amount %$w_i$
and processor speed, while communication time is the ratio between object
size (or output size) %$\delta_i$
and link bandwidth.

To assess performances, we study the
relative performance of each heuristic compared to the best solution
found by any heuristic.
This allows to compare the cost, in amount of resources used, of the different
heuristics. The relative performance for the heuristic $h$
is obtained by:
$\frac{1}{|runs|} \sum_{r=1}^{|runs|} a_h(r)$, where $a_h(r) = 0$ if
heuristic $h$ fails in run $r$ and $a_h(r) = \frac{cost_{best}(r)}{cost_{h}(r)}$.
$cost_{best}(r)$ is the best solution cost returned by one of the
heuristics for run $r$, and $cost_h(r)$ is the cost involved by the
solution proposed by heuristic~$h$. Note that in the definition
of the relative performance we do account for the case when a
heuristic fails on a given instance.
The number of runs is fixed to~$50$ in all experiments.
The complete
set of figures summarizing all experimental results is available on the web~\cite{figures.multiApp}.

\subsection{Results}
%\medskip
%\noindent
%\noindent
\subsubsection{Experiment 1: Number of Processors}
\begin{figure}[hb]
  \centering
  \subfigure[Successful runs, strategy 3]{
    \includegraphics[angle=270,width=0.49\textwidth]{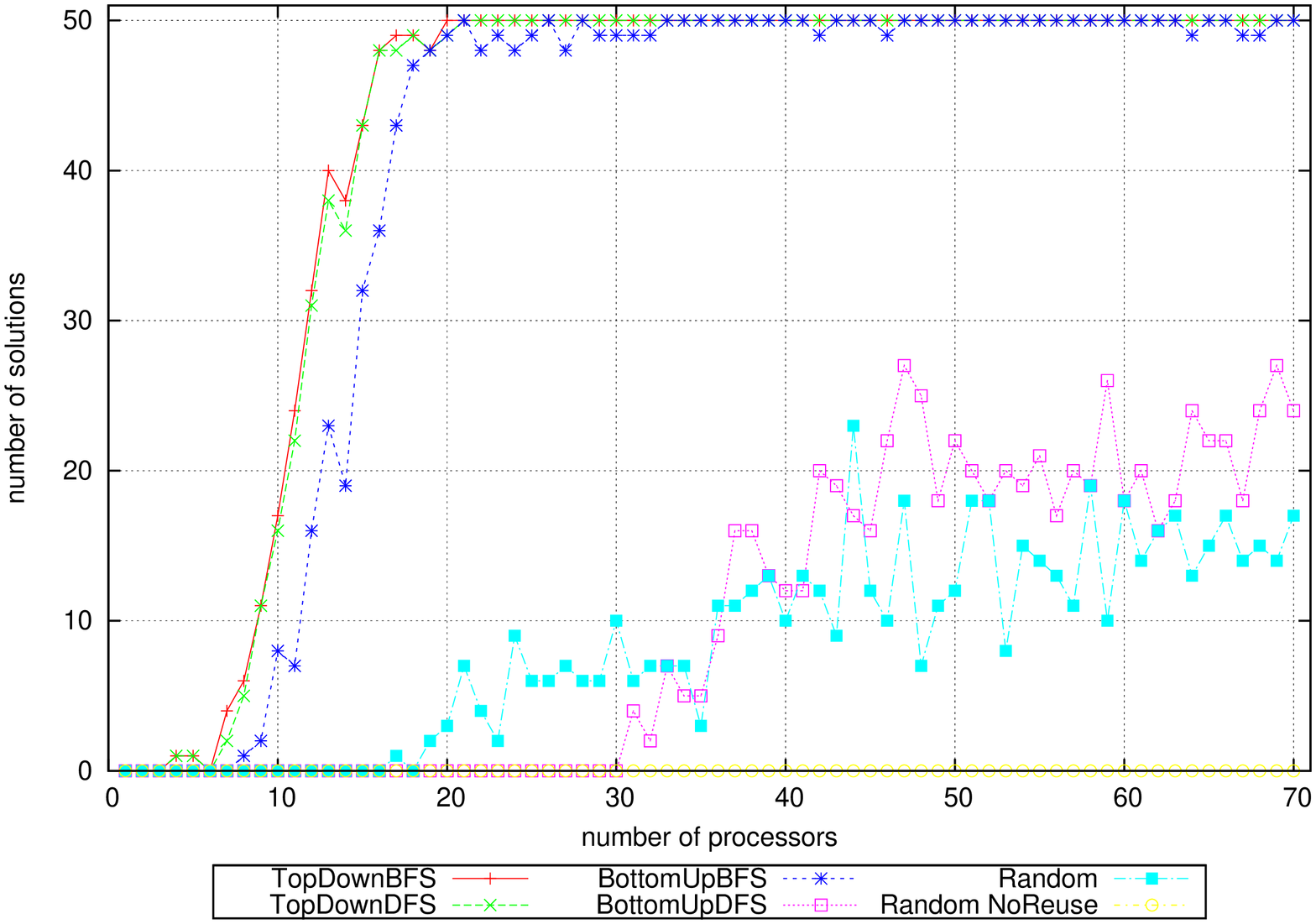}
    \label{fig.exp2.2.success}
  }$\quad$%
  \subfigure[Successful runs without reuse, strategy 3]{
    \includegraphics[angle=270,width=0.49\textwidth]{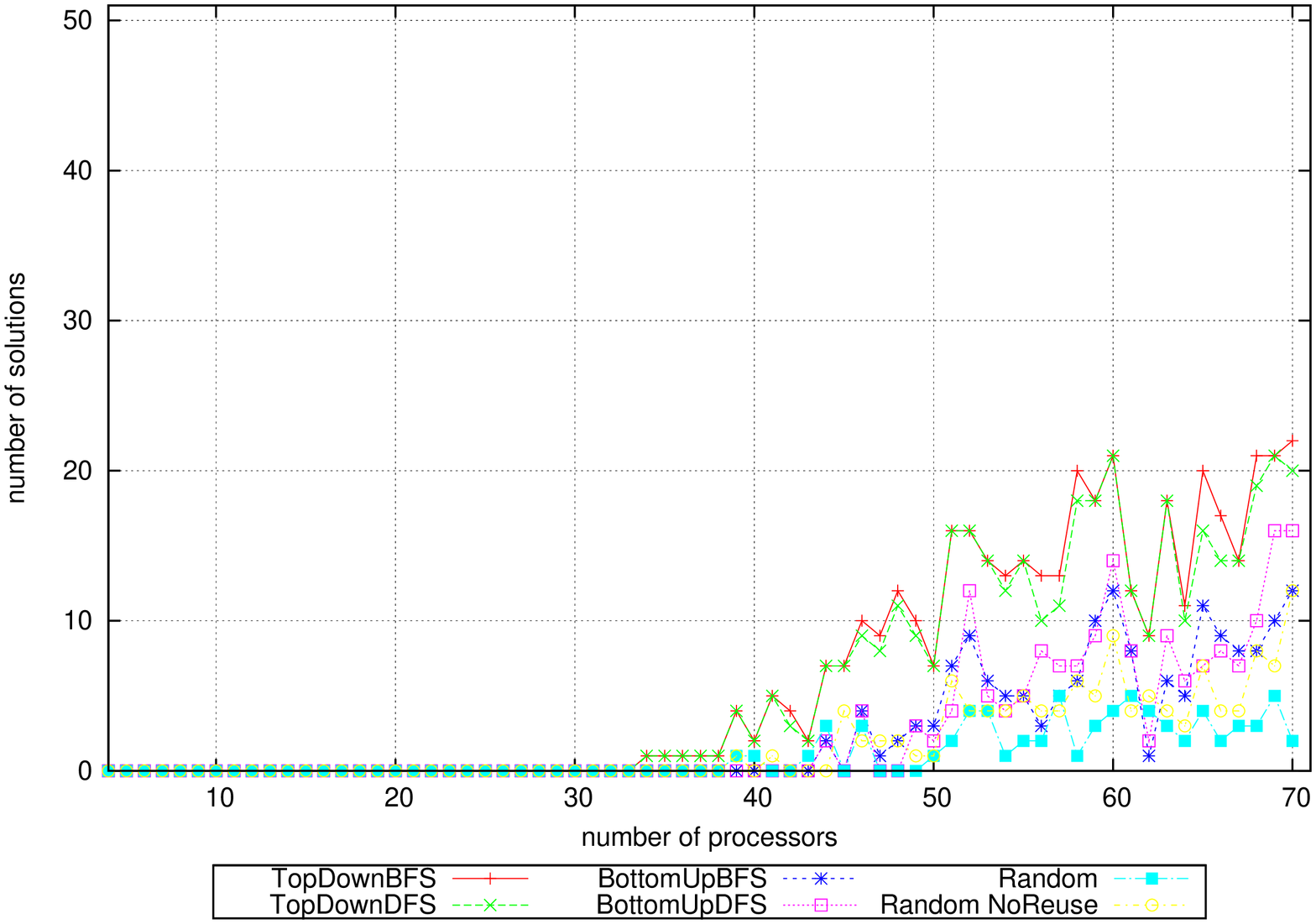}
    \label{fig.exp12.2.success}
  }
  \caption{Experiment 1: Increasing number of processors. Number of
    successful runs.}
\end{figure}

\begin{figure}
  \centering
  \subfigure[Relative performance, strategy 3]{
    \includegraphics[angle=270,width=0.49\textwidth]{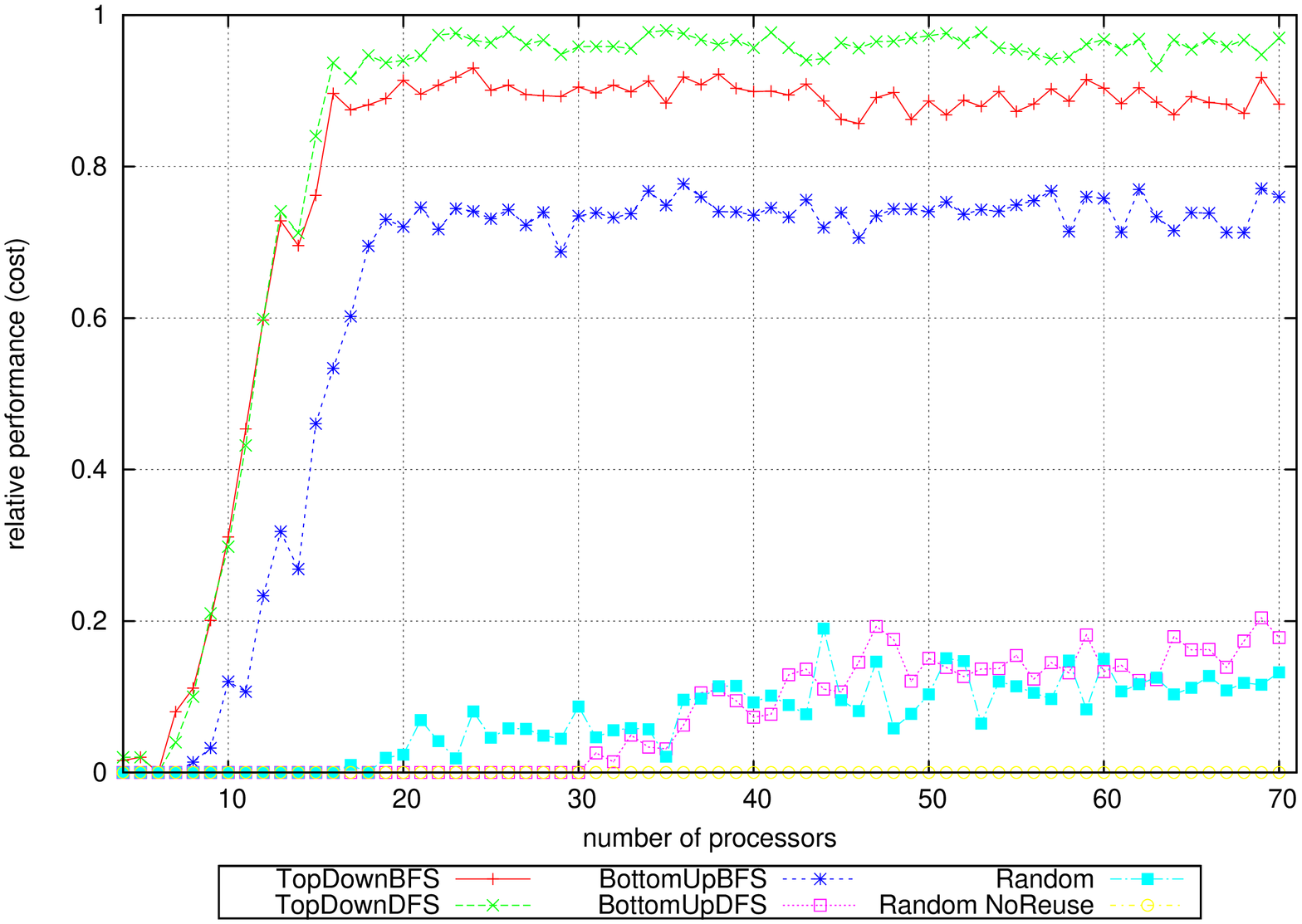}
    \label{fig.exp2.2.perf}
  }$\quad$%
  \subfigure[Relative performance, strategy 1]{
    \includegraphics[angle=270,width=0.49\textwidth]{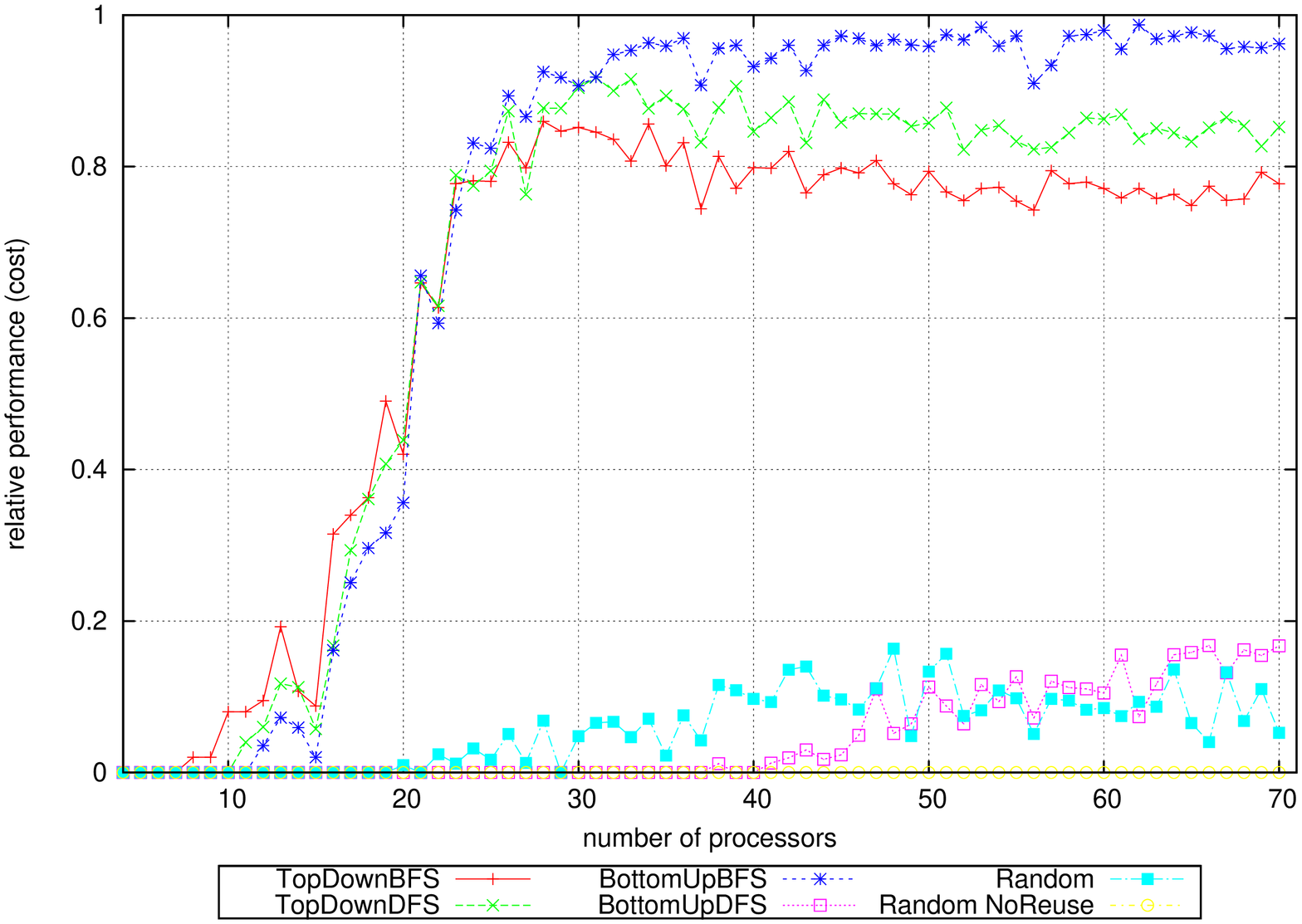}
    \label{fig.exp2.0.perf}
  }
  \caption{Experiment 1: Increasing number of processors. Relative performance.}
  \label{fix.exp2}
\end{figure}

%\label{sec.numProc}
%
In a first set of experiments, we test the influence of the number of available
processors, varying it from $1$ to~$70$.
%We work with 5 random application trees and
%vary the number of available processors from 1 to 70.
Figure~\ref{fig.exp2.2.success} shows the number of successes of the different
heuristics using selection strategy 3 (biggest remaining network card). Between
1 and 20 processors, the number of solutions
steeply increases for TopDownDFS, TopDownBFS and BottomUpBFS and
for higher numbers of processors  all three heuristics find solutions for most of
the 50 runs. BottomUpDFS finds solutions when more than 30
processors are available. Random already finds solutions
when only 20 processors are available, but for the runs with more than
30 processors, it finds fewer solutions than
BottomUpDFS. RandomNoReuse is not successful at all, it does not
find any solution. To summarize, TopDownBFS finds the most solutions,
shortly followed by TopDownDFS and BottomUpBFS.
Comparing the success rates of the different selection strategies, all
heuristics find the most solutions using strategy 3, followed by
strategy 4, strategy 2, and finally strategy 1. But the differences
are small. More interesting is the relative performance of the heuristics using
the different processor selection strategies in comparison to the
number of solutions.
Figure~\ref{fig.exp2.2.perf} shows the relative performance using
strategy~3. Comparing with
Figure~\ref{fig.exp2.0.perf}, we can conclude that for the same number of
successful runs, the performances of the heuristics significantly differ
according to
the selected processor selection strategy. Using strategy 3 (and also
strategies 2 and 4), TopDownDFS performs better than
TopDownBFS, which performs better than BottomUpBFS. However,
BottomUpBFS outperforms both TopDown heuristics when strategy
1 is used. The performance of BottomUpDFS and of the random ones
mirrors exactly the number of successful runs.
As for the heuristics without reuse of common subtrees, we see that
they do not find results until at least 35 processors are
available (strategy 3) or even 60 (strategy 2). Independently of the
processor selection strategy, both TopDown heuristics outperform all
other heuristics in success and performance, but the results are poor
(see Figure~\ref{fig.exp12.2.success}).

%% \medskip
%\noindent
\subsubsection{Experiment 2: Number of Applications}
\label{sec.numApp}
\begin{figure}[h]
  \centering
  \subfigure[Relative performance, strategy 1]{
    \includegraphics[angle=270,width=0.49\textwidth]{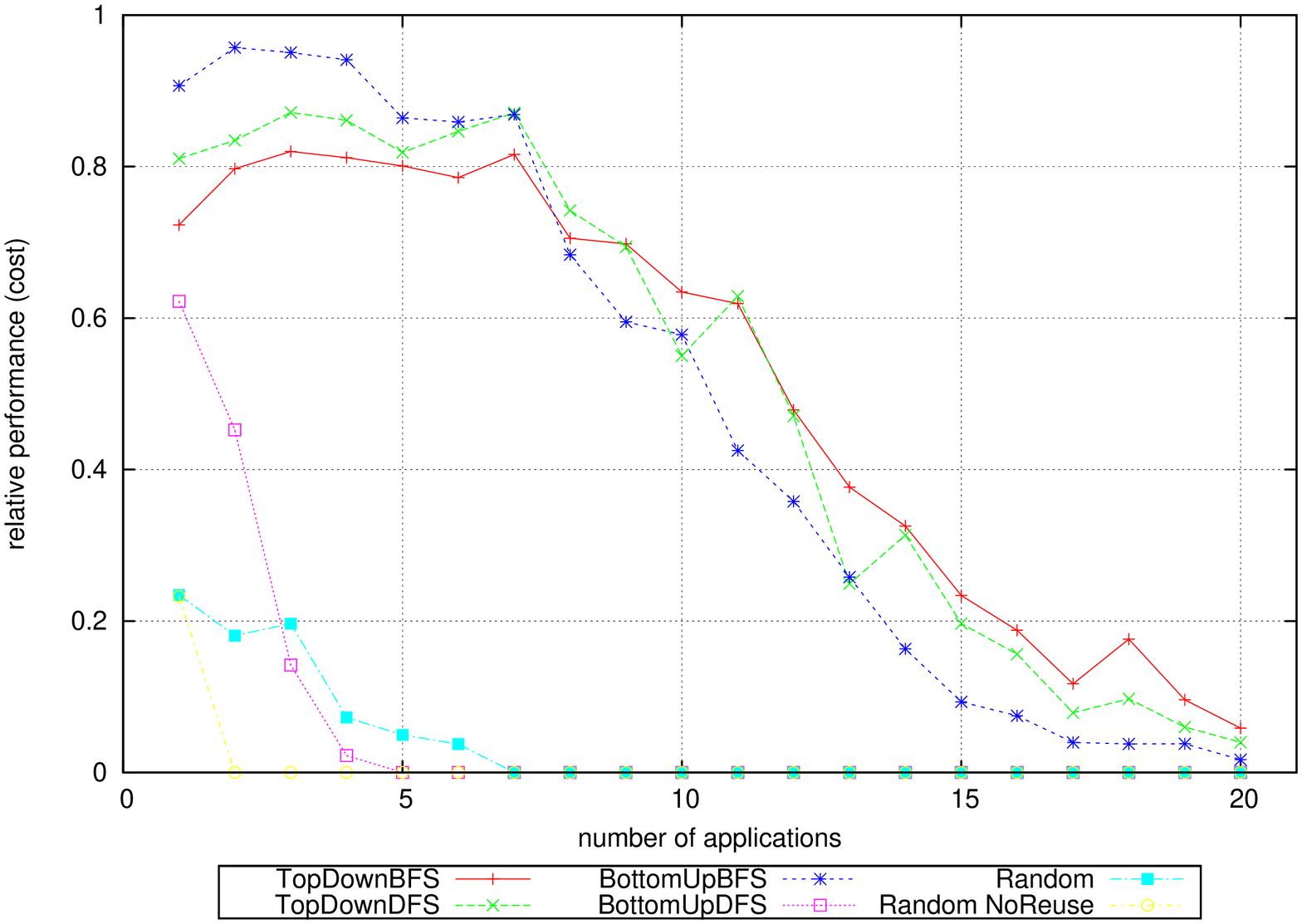}
    \label{fig.exp1.0.perf}
  }$\quad$%
  \subfigure[Relative performance, strategy 3]{
    \includegraphics[angle=270,width=0.49\textwidth]{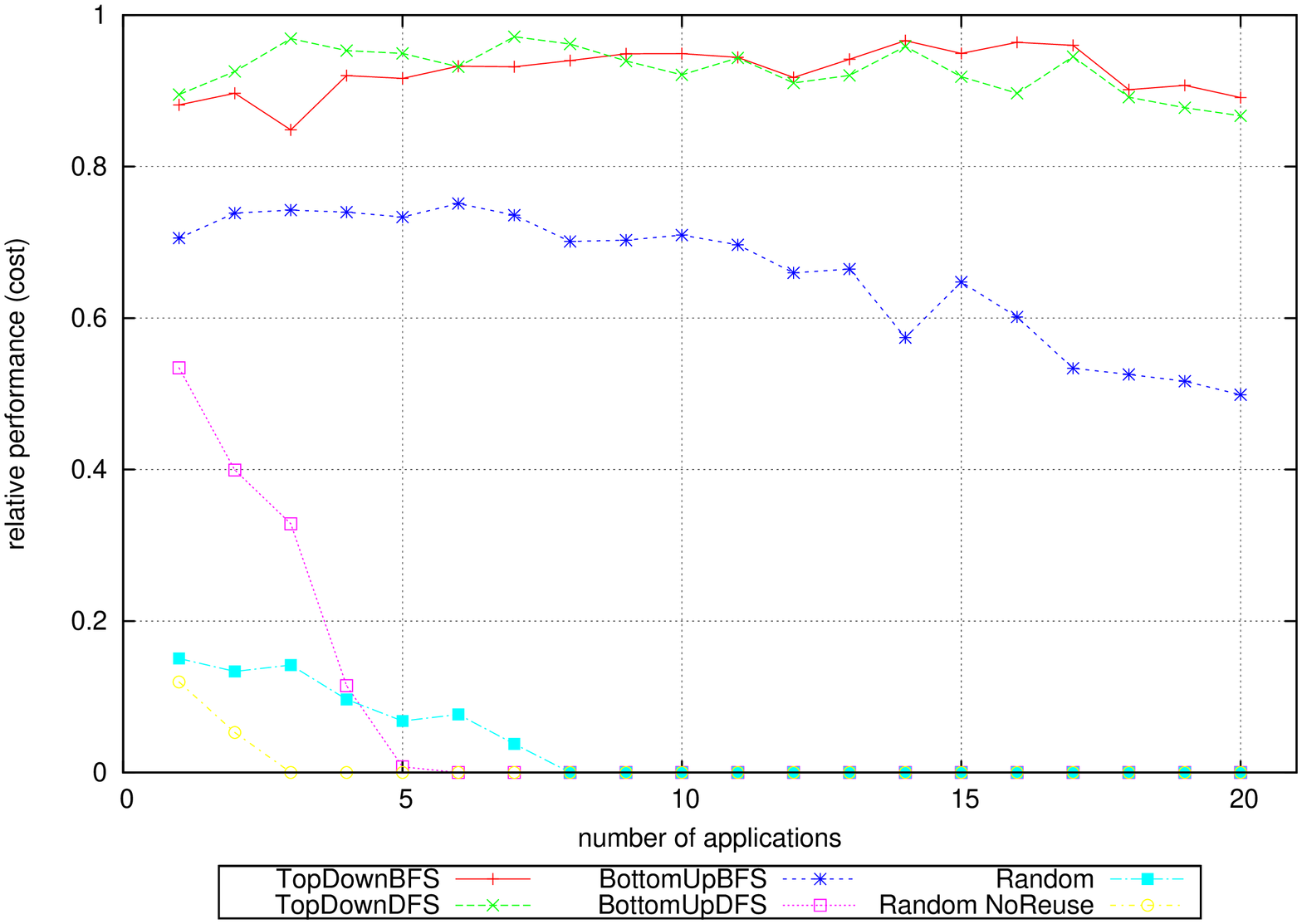}
    \label{fig.exp1.2.perf}
  }
  \caption{Experiment 2: Increasing number of applications.}
  \label{fig.exp1}
\end{figure}

In this set of experiments we vary
the number of applications, $\KK$.
As the number of application increases, all heuristics are
less successful with strategies 1 and 2 than with strategies
3 and 4, and 
relative
performance is poorer as well. Regardless of the strategy used, both
TopDown heuristics show a better relative performance than
BottomUpBFS, with the only exception using strategy 1 with a small
number of applications (Figure~\ref{fig.exp1.0.perf}). BottomUpDFS
and both random heuristics perform
poorly. For instance, BottomUpDFS only finds solutions with up to 4
applications. The best strategy seems to be strategy 3 in combination
with TopDownBFS for more than 10 applications and TopDownDFS for
less than 10 applications (see Figure~\ref{fig.exp1.2.perf}).

%% \medskip
%\noindent
\subsubsection{Experiment 3: Application Size}
\label{sec.appSize}
\begin{figure}[h]
  \centering
  \subfigure[Relative performance, strategy 1]{
    \includegraphics[angle=270,width=0.49\textwidth]{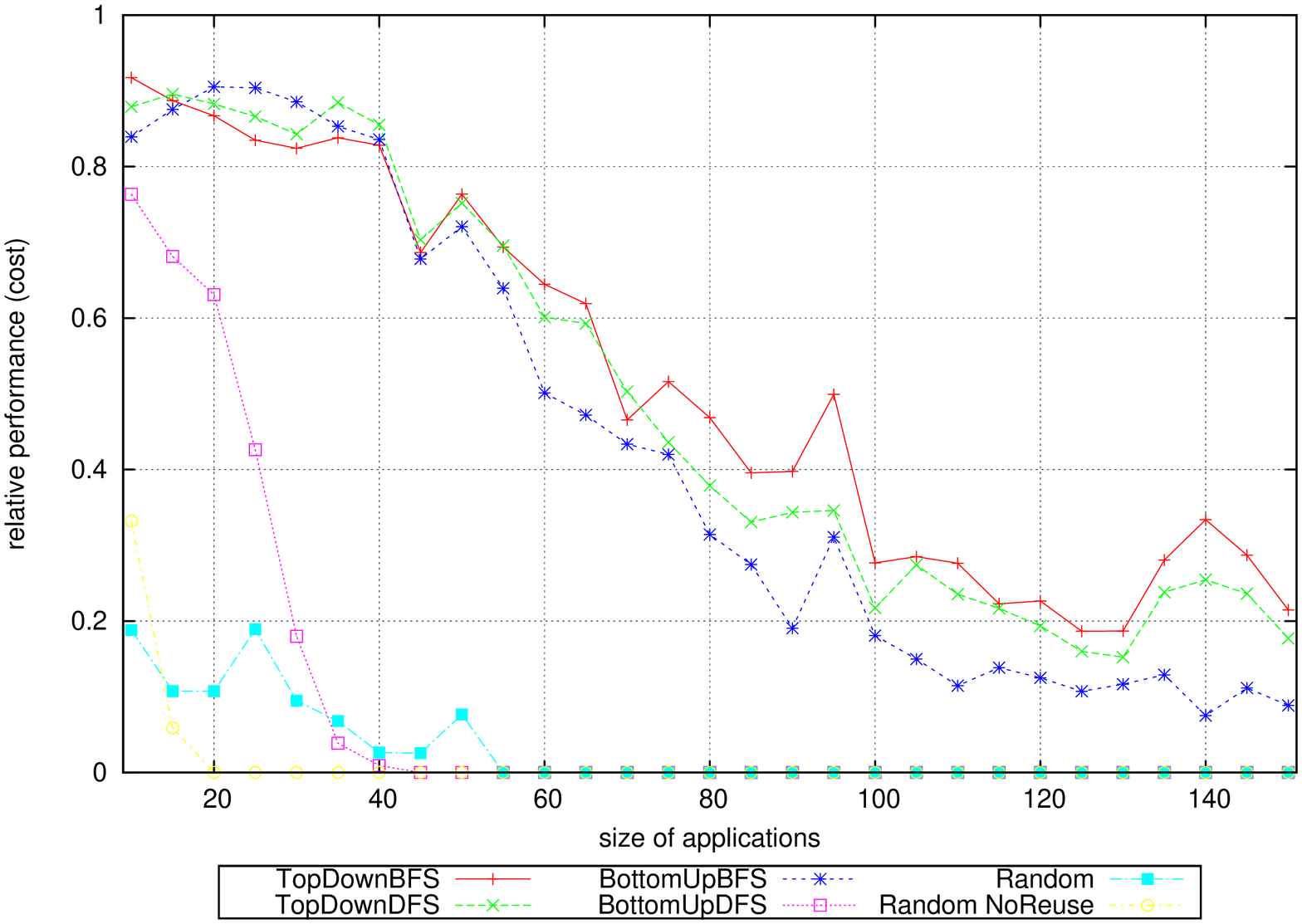}
    \label{fig.exp3.0.perf}
  }$\quad$%
  \subfigure[Relative performance, strategy 2]{
    \includegraphics[angle=270,width=0.49\textwidth]{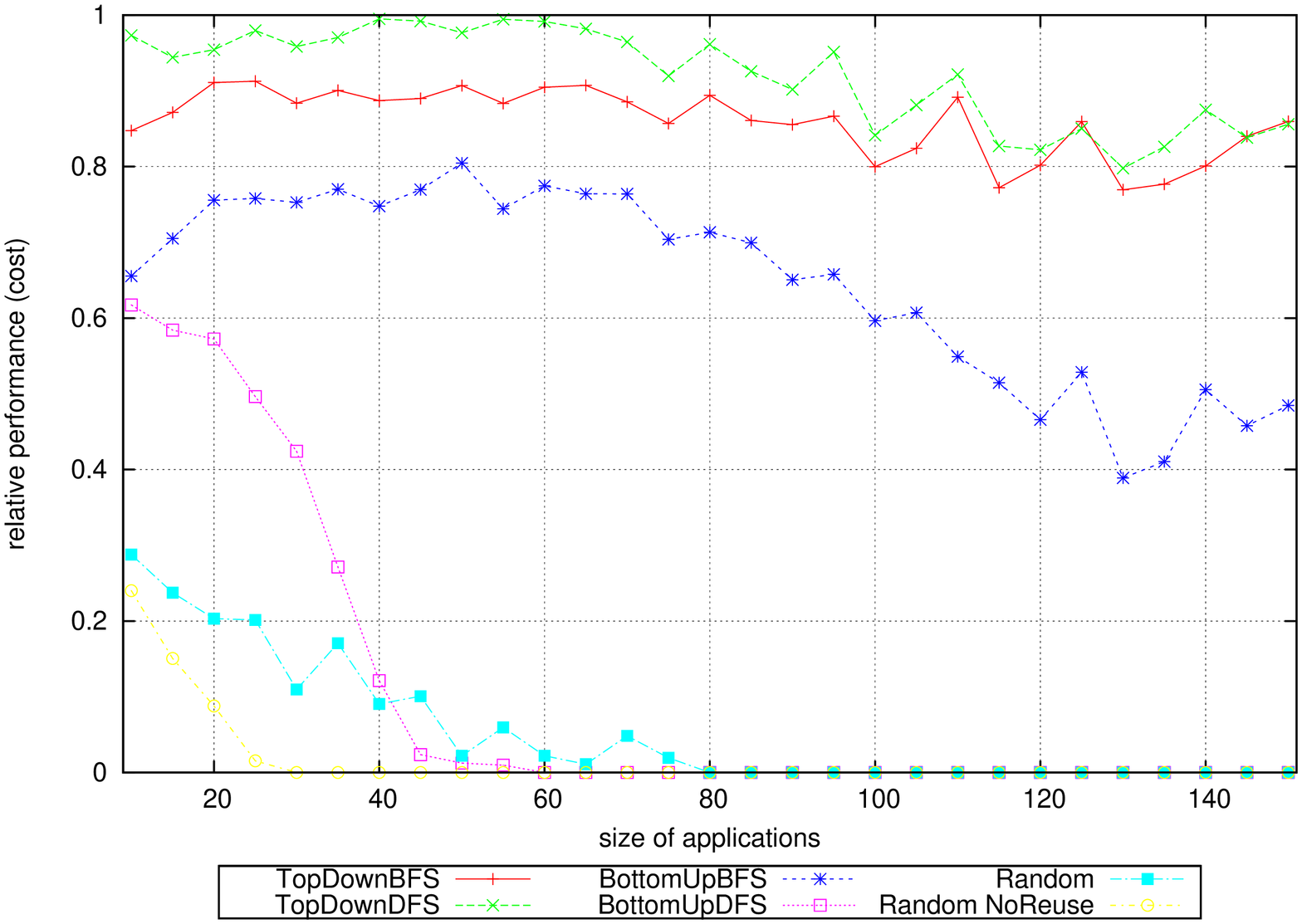}
    \label{fig.exp3.2.perf}
  }\caption{Experiment 3: Relative performance for increasing application sizes.}
\end{figure}

When increasing the application sizes, strategy 3 is the most robust.
Up to application sizes of 40 operators, the other strategies are
competitive, but for applications bigger than 40 operators both
TopDown heuristics and BottomUpBFS achieve the best relative
performance and find the most solutions. The success ranking of the three
heuristics is the same, independently of the strategy: TopDownBFS finds
more solutions than TopDownDFS, which, in turn, finds more solutions than
BottomUpBFS. RandomNoReuse finds solutions for applications with fewer 
than 20 operators, BottomUpBFS up to 40 operators and Random up to
50 operators, but the number of solutions from the latter is poor.
As far as relative performance is concerned, both TopDown heuristics achieve
the best results for application sizes bigger than 20 using strategy 3.
BottomUpDFS is competitive when using strategy 1 for
applications smaller than 40 operators (compare
Figures~\ref{fig.exp3.0.perf} and~\ref{fig.exp3.2.perf}).
As for the heuristics without reuse of common subtrees, they no longer
find results when application sizes
exceed 40 operators. TopDown heuristics perform better,
and the  best strategy is one of the two non-blocking ones (3 or 4).

%% \medskip
%\noindent
\subsubsection{Experiment 4: Communication-to-Computation Ratio (CCR)}
\label{sec.CCR}
\begin{figure}[h]
  \centering
  \subfigure[Successful runs, strategy 3]{
    \includegraphics[angle=270,width=0.49\textwidth]{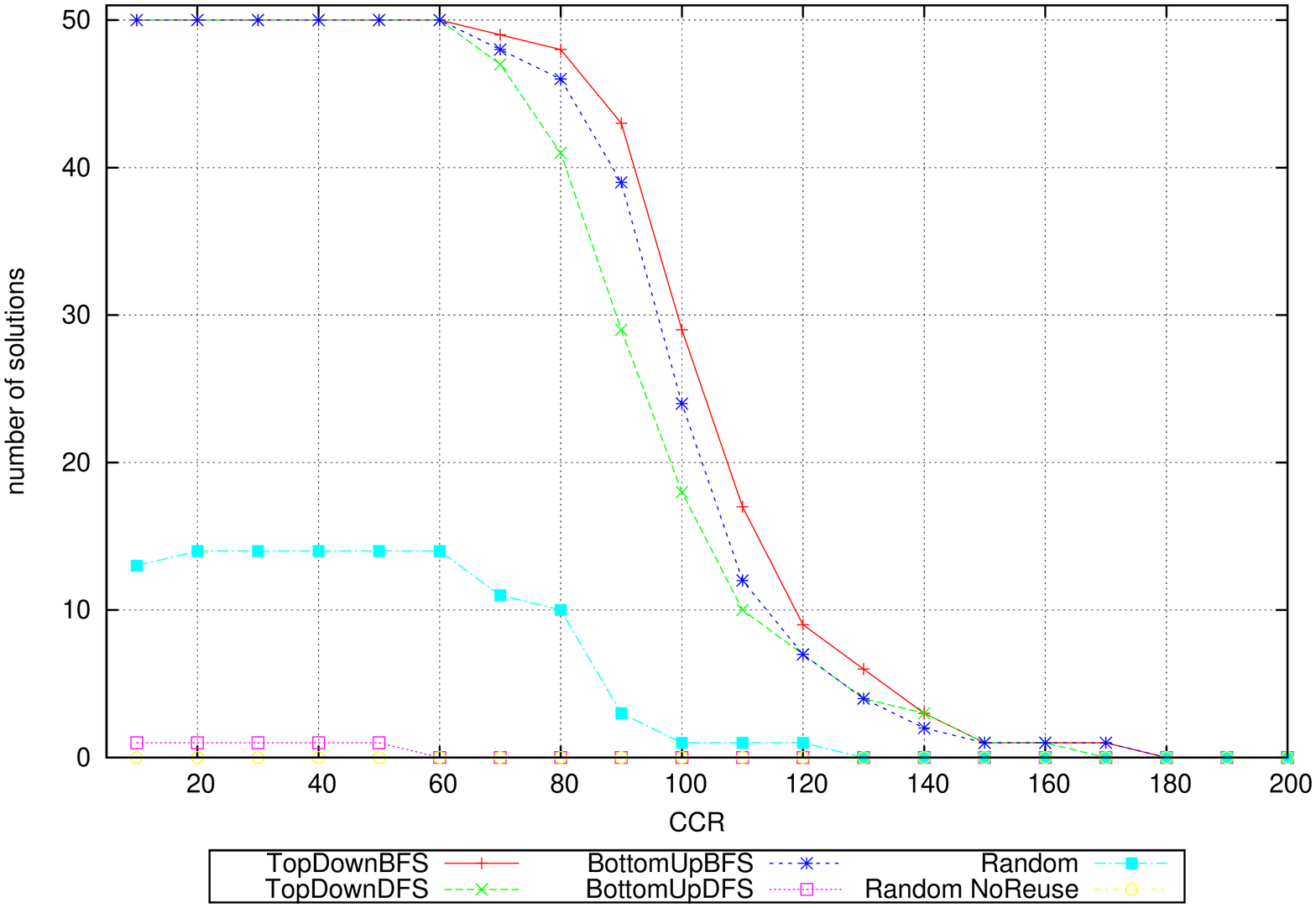}
    \label{fig.exp4.2.success}
  }$\quad$%
  \subfigure[Successful runs, strategy 2]{
    \includegraphics[angle=270,width=0.49\textwidth]{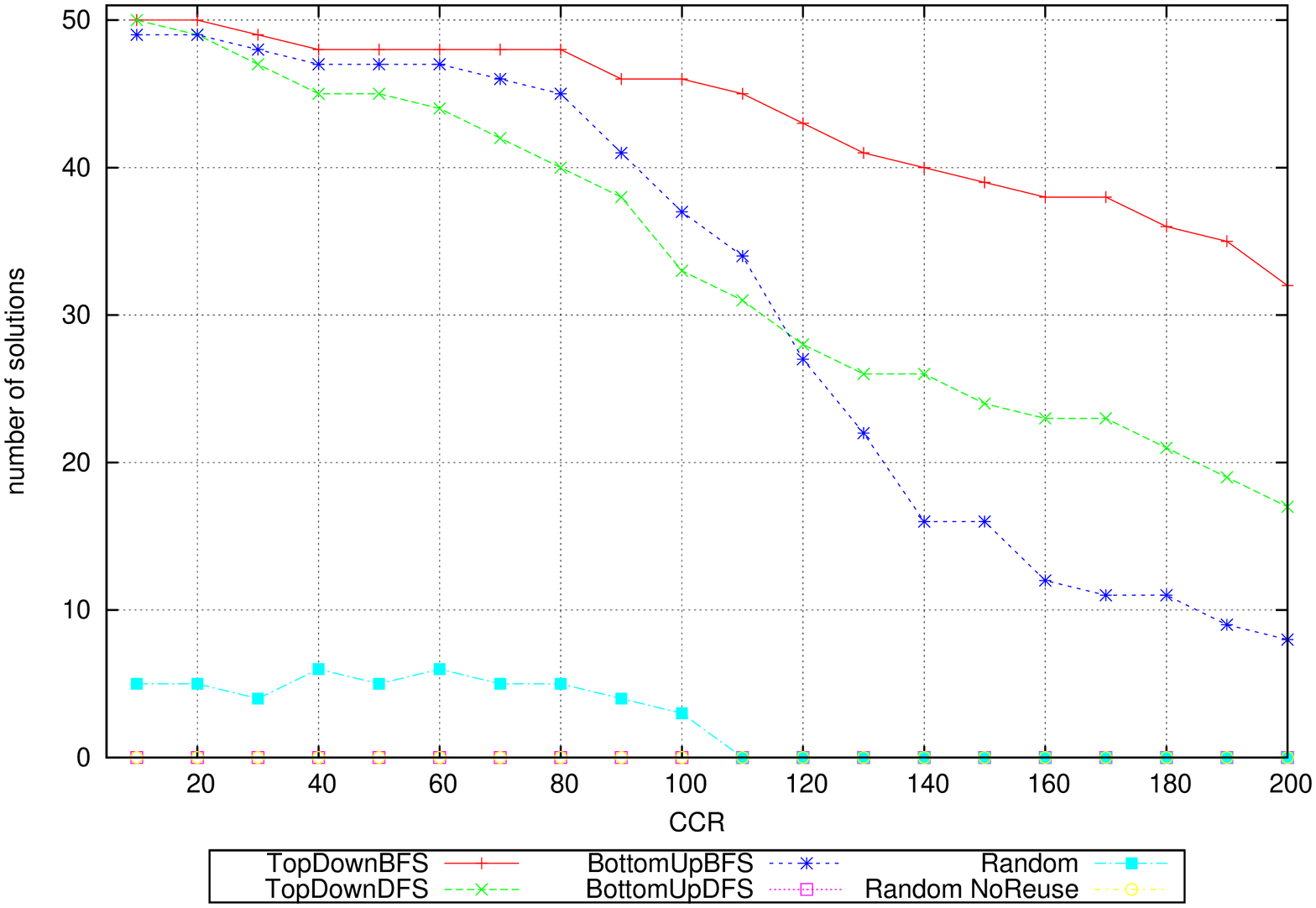}
    \label{fig.exp4.1.success}
  }
  \caption{Experiment 4: Communication-Computation Ratio
    CCR. Number of successful runs.}
\end{figure}

\begin{figure}
  \centering
  \subfigure[Relative performance, strategy 1]{
    \includegraphics[angle=270,width=0.49\textwidth]{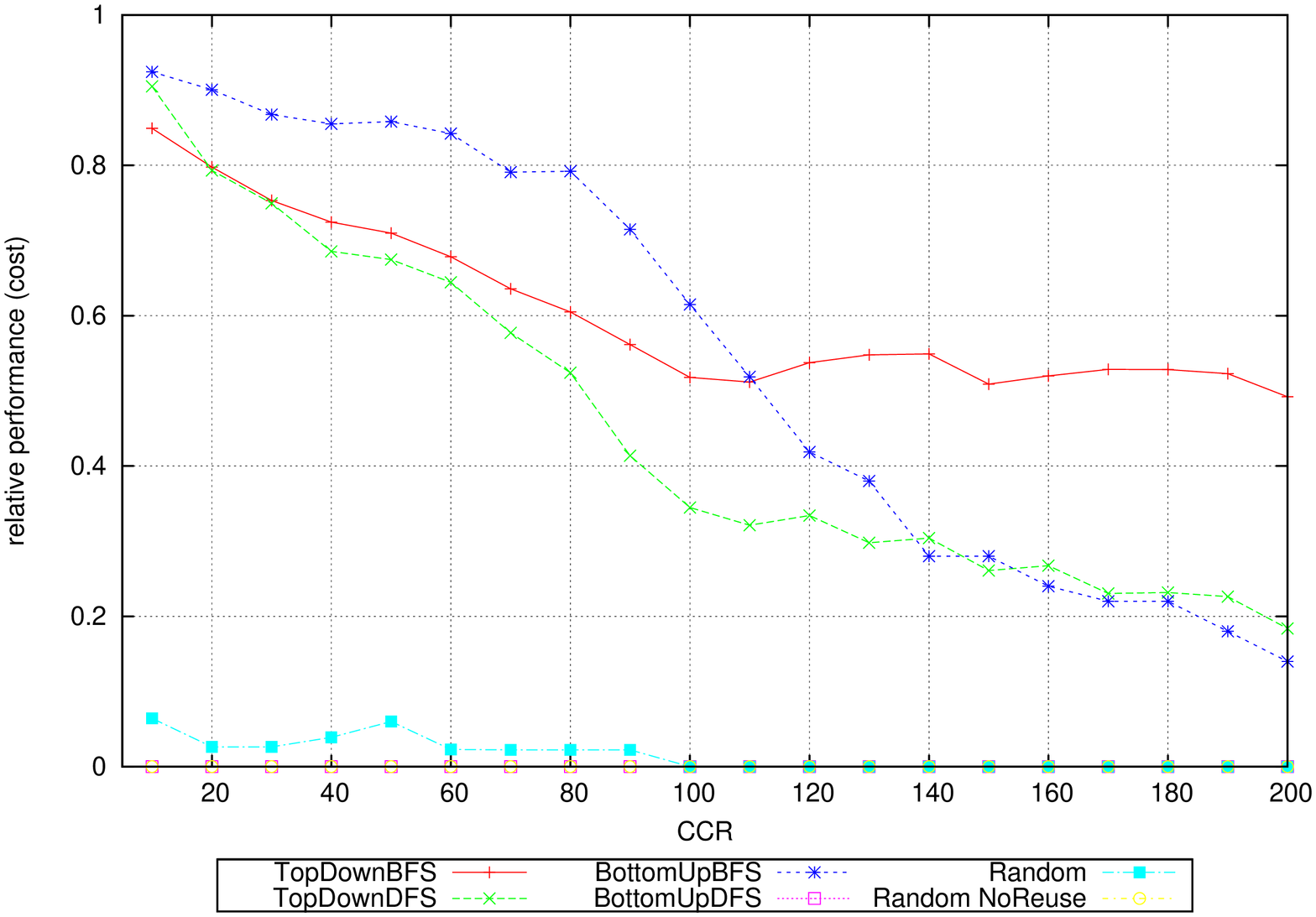}
    \label{fig.exp4.0.perf}
  }$\quad$%
  \subfigure[Relative performance, strategy 2]{
    \includegraphics[angle=270,width=0.49\textwidth]{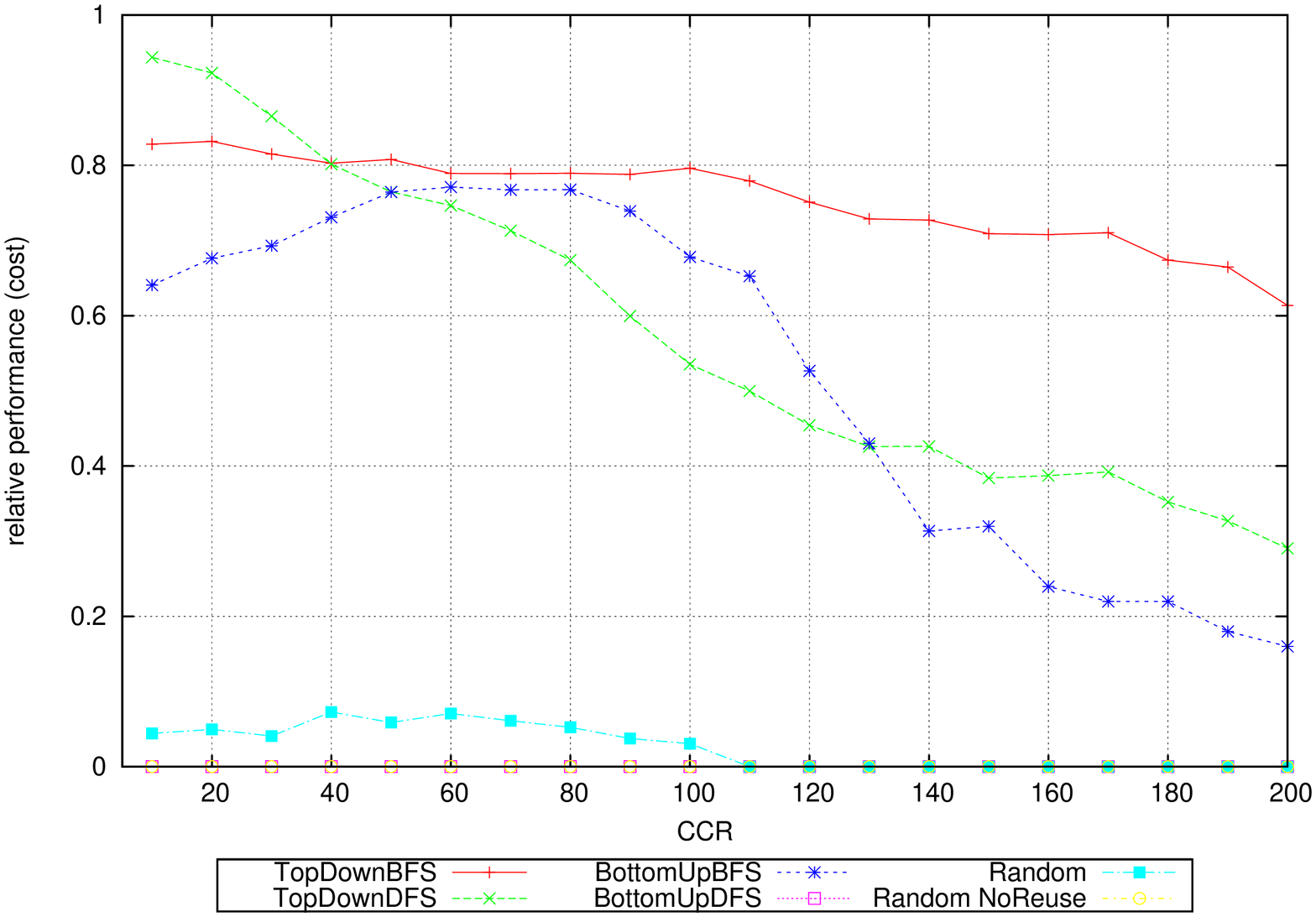}
    \label{fig.exp4.1.perf}
  }
  \caption{Experiment 4: Communication-Computation Ratio CCR. Relative
    performance.}
  \label{fig.exp4}
\end{figure}

For this experimental set we introduce a new parameter, the CCR, which
is the ratio between
the mean amount of communications and the mean amount of computations,
where the communications correspond to the output sizes of operators
($\delta_i$)
and the computations to the computational volume $w_i$ of the operators.
When increasing the CCR, strategies 3 and 4 react very
sensitively. As can be seen in Figure~\ref{fig.exp4.2.success},
TopDownBFS, TopDownDFS and BottomUpBFS have a 100\%
success rate for CCR $\leq 60$, but then the success decreases
drastically until no solution is found at
all for a CCR of 180 (using strategy 2, TopDownBFS still finds 32 solutions).
BottomUpDFS is largely outperformed by Random,
and RandomNoReuse fails completely.
In this experiment, strategy 2 seems to be the most successful processor
selection strategy (see Figure~\ref{fig.exp4.1.success}). TopDownBFS
achieves the best results, followed by BottomUpBFS for CCR $<120$,
and by TopDownDFS for CCR $> 120$.
Interestingly, the relative performances of the heuristics using the
different strategies do not directly mirror their success
rates. Compare Figures~\ref{fig.exp4.0.perf}
and~\ref{fig.exp4.1.perf}: BottomUpBFS finds fewer solutions using
strategy 1 than 2, but its relative performance using strategy 1 and
CCR smaller than 80
is better than when using strategy 2. Furthermore, TopDownBFS using
strategy 1 always
finds the most solutions of all heuristics, but its relative
performance is only the best when the CCR becomes bigger than
120. Also, TopDownDFS finds fewer solutions than
TopDownBFS and BottomUpBFS using strategy 2 and CCR$=30$, but its
relative performance is the best.

%% \medskip
%\noindent
\subsubsection{Experiment 5: Similarity of Applications}
\label{sec.similarity}
\begin{figure}[h]
  \centering
  \subfigure[Successful runs, strategy 2]{
    \includegraphics[angle=270,width=0.49\textwidth]{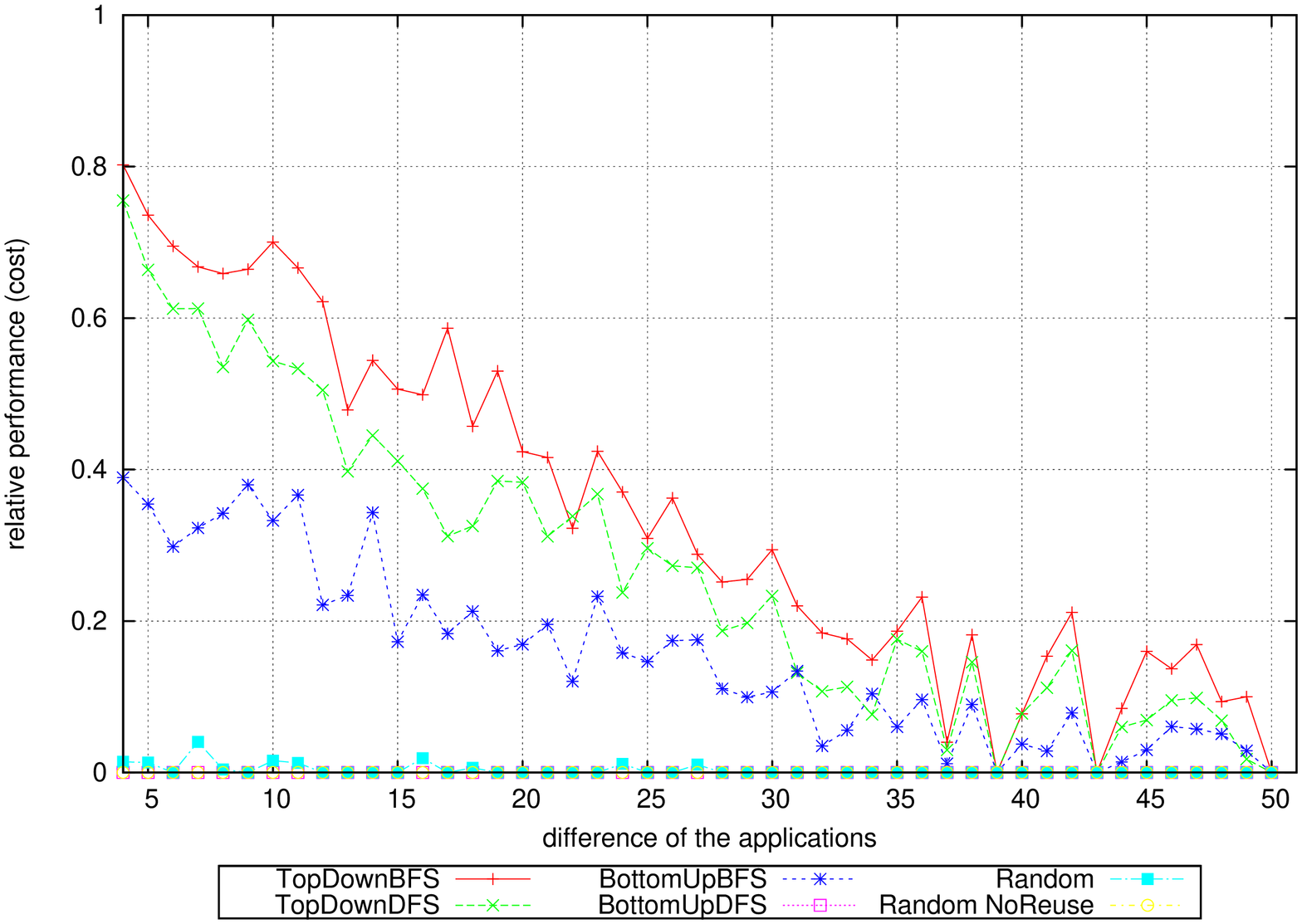}
    \label{fig.exp5.1.success}
  }$\quad$%
  \subfigure[Successful runs, strategy 3]{
    \includegraphics[angle=270,width=0.49\textwidth]{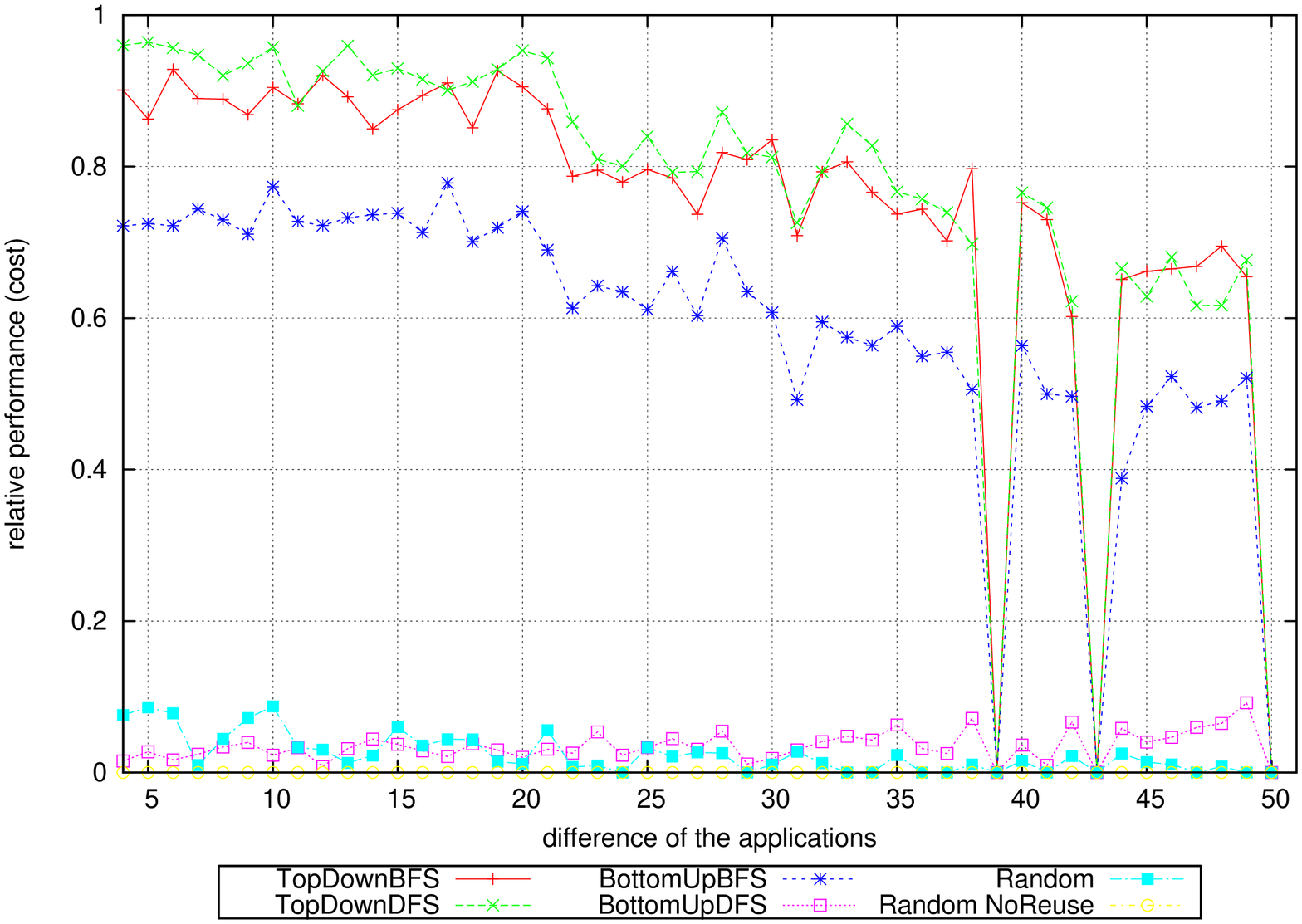}
    \label{fig.exp5.2.success}
  }
  \caption{Experiment 5: Similarity of applications.}
  \label{fig.exp5}
\end{figure}

In this last experiment, we use only two applications for each run and
the processing platform is smaller, consisting of only 10
processors. We study the influence on our heuristics when applications
are very similar or completely different. For this purpose we create
applications
 that differ in more and more operators.
Strategies 1 and 2 are more sensitive to application differences and
we observe the following ranking for the success of the heuristics:
strategy 3 > strategy 4 > strategies 1 and 2, which have similar success
rates (compare Figures~\ref{fig.exp5.1.success}
and~\ref{fig.exp5.2.success}.)
The ranking of the heuristics within the different strategies
is the same: TopDownBFS is the most successful, followed by
TopDownDFS and BottomUpBFS. BottomUpDFS and Random
keep the fourth place, while RandomNoReuse fails. TopDownBFS has the
best relative performance using the blocking strategies, whereas in
the non-blocking cases TopDownDFS achieves the best
results, which is important as its success rate is slightly poorer.
BottomUpBFS always ranks at the third position.

%% \medskip
%\noindent
\subsubsection{Summary of Experiments}
\label{sec.summary}
Our results show that a random approach for multiple applications is
not feasible. Neglecting the possibility to reuse results from common
subtrees dramatically limits the success rate and also the quality of the
solution in terms of cost. The TopDown approach turns out to be the best,
whereupon in most cases BFS traversal achieves the best
result. The BottomUp approach is only competitive using a BFS
traversal. The DFS traversal seems  unable to reuse results efficiently (it
often finds itself with no bandwidth left to perform necessary communications.)
Furthermore we see a strong dependency of the processor selection strategy
on solution quality. The blocking strategies outperform the non-blocking
strategies when the CCR is large. In the other cases, TopDownBFS in
combination with strategy 3 proves to be a solid combination.

%\section{Related Work}
%\label{sec.related}
%\remark{May be sufficient to keep related work in the introduction}

\section{Conclusion}
\label{sec.conclusion}
In this paper, we have studied the operator mapping problem of multiple
concurrent in-network stream-processing applications onto a collection of
heterogeneous processors. These stream-processing applications come as a set of
operator trees, that have to continuously download basic objects at
different sites of the network and at the same time have to process
this data to produce some final result.
We have considered the problem under a non-constructive scenario, in which 
a fixed set of computation and communication resources is available and the goal is %either
%to use the less resources as possible or
to minimize a cost
function. Four different optimization problems were identified. All
are NP-hard but can be formalized as integer linear programs.
On the practical side we focused on one of the optimization problems, for
which we designed a set of polynomial-time heuristics. We evaluated these
heuristics via extensive simulations, and our experiments showed the
importance of node reuse across applications. Reusing nodes leads to an
important number of additional solutions, and also the quality of the
solutions improves considerably. We concluded that top-down traversals of
the application trees is more efficient than bottom-up approaches, and in
particular the combination of a top-down traversal with a breadth-first
search (i.e., our heuristic TopDownBFS) achieved good results across the
board.

As future work, we could develop heuristics for the other optimization
problems defined in Section~\ref{sec.prob.obj}. We could also envision a more
general cost function $\calc_{i,u}$ (time required to compute operator~$i$
onto processor~$u$), in order to express even more heterogeneity. This
would lead to the design of more sophisticated heuristics. Also, we believe
it would be interesting to add a storage cost for objects downloaded onto
processors, which could lead to new objective functions.  Finally, we could
address more complicated scenarios with many (conflicting) relevant
criteria to consider simultaneously, some related to performance
(throughput, response time), some related to safety (replicating some
computations for more reliability), and some related to environmental costs
(resource costs, energy consumption).

%\newpage
%%\pagebreak
\bibliographystyle{plain}
\bibliography{biblio}

\begin{thebibliography}{10}

\bibitem{figures.multiApp}
{Diagrams of all experiments}.
\newblock http://graal.ens-lyon.fr/$\sim$vsonigo/code/query-multiapp/diagrams/.

\bibitem{code.multiApp}
{Source Code for the Heuristics}.
\newblock http://graal.ens-lyon.fr/$\sim$vsonigo/code/query-multiapp/.

\bibitem{abadi2005design}
Daniel~J Abadi, Yanif Ahmad, Magdalena Balazinska, Ugur Cetintemel, Mitch
  Cherniack, Jeong-Hyon Hwang, Wolfgang Lindner, Anurag~S Maskey, Alexander
  Rasin, Esther Ryvkina, Nesime Tatbul, Ying Xing, and Stan Zdonik.
\newblock {The Design of the Borealis Stream Processing Engine}.
\newblock In {\em Second Biennial Conference on Innovative Data Systems
  Research (CIDR 2005)}, Asilomar, CA, January 2005.

\bibitem{Babu_SIGMODRECORD_2001}
S.~Babu and J.~Widom.
\newblock {Continuous Queries over Data Streams}.
\newblock {\em SIGMOD Record}, 30(3), 2001.

\bibitem{badcock_VLDB_2004}
B.~Badcock, S.~Babu, M.~Datar, R.~Motwani, and J.~Widom.
\newblock {Models and issues in data stream systems}.
\newblock In {\em Proceedings of the Intl. Conf. on Very Large Data Bases},
  pages 456--467, 2004.

\bibitem{RR2008-20}
Anne Benoit, Henri Casanova, Veronika Rehn-Sonigo, and Yves Robert.
\newblock {Resource Allocation Strategies for Constructive In-Network Stream
  Processing}.
\newblock Research Report 2008-20, LIP, ENS Lyon, France, June 2008.

\bibitem{APDCM_09}
Anne Benoit, Henri Casanova, Veronika Rehn-Sonigo, and Yves Robert.
\newblock {Resource Allocation Strategies for Constructive In-Network Stream
  Processing}.
\newblock In {\em Proceedings of APDCM'09, the 11th Workshop on Advances in
  Parallel and Distributed Computational Models}. IEEE, 2009.

\bibitem{bonnet_CMDB_2001}
P.~Bonnet, J.~Gehrke, and P.~Seshadri.
\newblock {Towards sensor database systems}.
\newblock In {\em Proceedings of the Conference on Mobile Data Management},
  2001.

\bibitem{NiagaraCQ}
J.~Chen, D.J. DeWitt, F.~Tian, and Y.~Wang.
\newblock {NiagaraCQ: A scalable continuous query system for internet
  databases}.
\newblock In {\em Proceedings of the SIGMOD Intl. Conf. on Management of Data},
  pages 379--390, 2000.

\bibitem{chen02design}
Jianjun Chen, David~J. DeWitt, and Jeffrey~F. Naughton.
\newblock {Design and Evaluation of Alternative Selection Placement Strategies
  in Optimizing Continuous Queries}.
\newblock In {\em {Proceedings of ICDE}}, 2002.

\bibitem{gates}
Liang Chen, K.~Reddy, and G.~Agrawal.
\newblock {GATES: a grid-based middleware for processing distributed data
  streams}.
\newblock {\em High performance Distributed Computing, 2004. Proceedings. 13th
  IEEE International Symposium on}, pages 192--201, 4-6 June 2004.

\bibitem{medusa}
M.~Cherniack, H.~Balakrishnan, M.~Balazinska, D.~Carney, U.~Cetintemel,
  Y.~Xing, and S.~Zdonik.
\newblock Scalable distributed stream processing.
\newblock In {\em Proc. of the CIDR Conf.}, January 2003.

\bibitem{cooke_USENIX_2006}
E.~Cooke, R.~Mortier, A.~Donnelly, P.~Barham, and R.~Isaacs.
\newblock {Reclaiming Network-wide Visibility Using Ubiquitous End System
  Monitors}.
\newblock In {\em Proceedings of the USENIX Annual Technical Conference}, 2006.

\bibitem{cranor_ICMD_2002}
C.~Cranor, Y.~Gao, T.~Johnson, V.~Shkapenyuk, and O.~Spatscheck.
\newblock {Gigascope: high-performance network monitoring with an SQL
  interface}.
\newblock In {\em Proceedings of the ACM SIGMOD International Conference on
  Management of Data}, pages 623--633, 2002.

\bibitem{GareyJohnson}
M.~R. Garey and D.~S. Johnson.
\newblock {\em Computers and Intractability, a Guide to the Theory of
  {NP}-Completeness}.
\newblock W.H. Freeman and Company, 1979.

\bibitem{bohong04}
B.~Hong and V.K. Prasanna.
\newblock Distributed adaptive task allocation in heterogeneous computing
  environments to maximize throughput.
\newblock In {\em International Parallel and Distributed Processing Symposium
  {IPDPS'2004}}. IEEE Computer Society Press, 2004.

\bibitem{pier}
Ryan Huebsch, Joseph~M. Hellerstein, Nick~Lanham Boon, Thau Loo, Scott Shenker,
  and Ion Stoica.
\newblock {Querying the Internet with PIER}, September 2003.

\bibitem{ioannidis96query}
Yannis~E. Ioannidis.
\newblock Query optimization.
\newblock {\em ACM Computing Surveys}, 28(1):121--123, 1996.

\bibitem{kramer_COMAD05}
J.~Kr\"ame and B.~Seeger.
\newblock {A Temporal Foundation for Continuous Queries over Data streams}.
\newblock In {\em Proceedings of the Intl. Conf. on Management of Data}, pages
  70--82, 2005.

\bibitem{liu_tkde1999}
L.~Liu, C.~Pu, and W.~Tang.
\newblock {Continual Queries for Internet Scale Event-Driven Information
  Delivery}.
\newblock {\em IEEE Transactions on Knowledge and Data Engineering},
  11(4):610--628, 1999.

\bibitem{MORTAR}
D.~Logothetis and K.~Yocum.
\newblock {Wide-Scale Data Stream Management}.
\newblock In {\em Proceedings of the USENIX Annual Technical Conference}, 2008.

\bibitem{madden_ICMD_2003}
S.~Madden, M.~Franklin, J.~Hellerstein, and W.~Hong.
\newblock {The design of an acquisitional query processor for sensor networks}.
\newblock In {\em Proceedings of the 2003 ACM SIGMOD Intl. Conf. on Management
  of Data}, pages 491--502, 2003.

\bibitem{nath-irisnet}
Suman Nath, Amol Deshpande, Yan Ke, Phillip~B. Gibbons, Brad Karp, and
  Srinivasan Seshan.
\newblock {IrisNet: An Architecture for Internet-scale Sensing Services}.

\bibitem{Pandit06}
Vinayaka Pandit and Huibo Ji.
\newblock Efficient in-network evaluation of multiple queries.
\newblock In {\em HiPC}, pages 205--216, 2006.

\bibitem{Pietzuch_ICDE06}
P.~Pietzuch, J.~Leflie, J.~Shneidman, M.~Roussopoulos, M.~Welsh, and
  M.~Seltzer.
\newblock {Network-Aware Operator Placement for Stream-Processing Systems}.
\newblock In {\em Proceedings of the 22nd International Conference on Data
  Engineering (ICDE'06)}, pages 49--60, 2006.

\bibitem{Plale_TPDD_2003}
B.~Plale and K.~Schwan.
\newblock {Dynamic Querying of Streaming Data with the dQUOB System}.
\newblock {\em IEEE Transactions on Parallel and Distributed Systems},
  14(4):422--432, 2003.

\bibitem{srivastava_PODS2005}
U.~Srivastava, K.~Munagala, and J.~Widom.
\newblock {Operator Placement for In-Network Stream Query Processing}.
\newblock In {\em Proceedings of the 24th ACM Intl. Conf. on Principles of
  Database Systems}, pages 250--258, 2005.

\bibitem{vanRennesse_IPTPS_2002}
R.~van Rennesse, K.~Birman, D.~Dumitriu, and W.~Vogels.
\newblock {Scalable Management and Data Mining Using Astrolabe}.
\newblock In {\em Proceedings from the First International Workshop on
  Peer-to-Peer Systems}, pages 280--294, 2002.

\end{thebibliography}

\end{document}